\documentclass[english,aps,prb,superscriptaddress]{revtex4-1}
\usepackage[T1]{fontenc}
\usepackage[latin9]{inputenc}
\usepackage{xcolor}
\usepackage{textcomp}
\usepackage{amstext}
\usepackage{graphicx}
\usepackage{subscript}

\makeatletter

\providecommand{\tabularnewline}{\\}

\usepackage{braket}

\@ifundefined{showcaptionsetup}{}{%
 \PassOptionsToPackage{caption=false}{subfig}}
\usepackage{subfig}
\makeatother

\usepackage{babel}
\begin{document}
\title{Influence of Edge Functionalization on Electronic and Optical Properties
of Armchair Phosphorene Nanoribbons: a First-Principles Study}
\author{Pritam Bhattacharyya}
\email{pritambhattacharyya01@gmail.com}

\affiliation{Department of Physics, Indian Institute of Technology Bombay, Powai,
Mumbai 400076, India}
\author{Rupesh Chaudhari}
\affiliation{Department of Physics, Indian Institute of Technology Bombay, Powai,
Mumbai 400076, India}
\author{Naresh Alaal}
\email{nareshkdnr@gmail.com}

\affiliation{Physical Sciences and Engineering Division, King Abdullah University
of Science and Technology, Thuwal 23955-6900, Saudi Arabia}
\author{Tushar Rana}
\affiliation{Department of Physics and Nanotechnology, SRM Institute of Science
and Technology,SRM Nagar, Kattankulathur - 603203 (Tamil Nadu), India}
\author{Alok Shukla}
\email{shukla@phy.iitb.ac.in}

\affiliation{Department of Physics, Indian Institute of Technology Bombay, Powai,
Mumbai 400076, India}
\begin{abstract}
In this work, we present a systematic first-principles density-functional
theory based study of geometry, electronic structure, and optical
properties of armchair phosphorene nanoribbons (APNRs), with the aim
of understanding the influence of edge passivation. Ribbons of width
ranging from 0.33 nm to 3.8 nm were considered, with their edges functionalized
with the groups H, OH, F, Cl, S, and Se. The geometries of various
APNRs were optimized, and the stability was checked by calculating
their formation energies. Using the relaxed geometries, calculations
of their band structure and optical properties were performed. Pristine
APNRs, as expected, exhibit significant edge reconstruction, rendering
them indirect band gap semiconductors, except for one width {\normalsize{}($N=5$,
where $N$ is the width parameter)} for which a direct band gap is
observed. The edge passivated APNRs are found to be direct band gap
semiconductors, with the band gap at the $\Gamma$-point, for all
the functional groups\textcolor{red}{{} }considered in this work. To
obtain accurate estimates of band gaps, calculations were also performed
using HSE06 hybrid functional for several APNRs. Our calculations
reveal that functional groups have significant influence on the band
gaps and optical properties of narrower APNRs. For wider passivated
ribbons, with the increasing ribbon widths, the gaps converge to almost
the same value, irrespective of the group. We also performed calculations
including the spin-orbit coupling (SOC) for hydrogen passivated APNRs
with $N=5$ and 11. We found that SOC has no significant influence
on the band structure of the studied APNRs. However, for the broader
APNR, a lowering of peak intensities was observed in the optical absorption
spectrum beyond 5 eV. 
\end{abstract}
\maketitle

\section{Introduction}

For the past few decades, there has been tremendous amount of interest
in reduced-dimensional systems, in general, and two-dimensional (2D)
materials, in particular.\citep{2D_review_1,2D_review_2,2D_review_3,2D_review_4,2D_review_5}
Not only are 2D materials interesting from the basic science point
of view, they also offer easier tuning of their electronic properties,
as compared to their bulk counterparts. The tailoring of their electronic
properties is crucial to make them suitable for applications in electronics
and optics. Phosphorene not only has a finite direct band gap, but
also has high carrier mobility, and in-plane anisotropy. This 2D-material
offers many potential applications not only in transport-based electronic
and spintronic devices, but also in sensors, information storage,
and optoelectronic devices.\citep{Akhtar_et_al} After the successful
exfoliation of phosphorene from bulk black phosphorus,\citep{Li2014_et_al,Liu_et_al,Woomer_et_al,Brent_et_al,Zhinan_et_al,Ambrosi_et_al}
it has been a subject of extensive experimental as well as theoretical
research.\citep{Liu_et_al,Likai_et_al,Koenig_et_al,Dai_et_al,Weinan_1_et_al,Weinan_2_et_al}
The band gap of phosphorene can be further tailored by manipulating
the number of layers, in-layer strain,\citep{Liu_et_al} forming heterostructures
such as nanoribbons and quantum dots, as well as by chemical means
such as edge passivation.

In this work, we present a systematic first-principles density functional
theory (DFT) based study of the geometry, electronic structure, and
optical properties of armchair-type phosphorene nanoribbons (APNRs),
with the aim of understanding the influence of edge-passivation on
them. Numerous theoretical studies of phosphorene and its heterostructures
have been performed over last few years, therefore, it is difficult
to cite all of them. However, below we review the most relevant theoretical
studies of the electronic structure and related properties of APNRs,
which have been studied mainly using two methodologies: (a) semi-empirical
tight-binding model, and (b) first-principles DFT.

Using the tight-binding model Sisakht and coworkers\citep{Sisakht_et_al}
studied the scaling laws in phosphorene nanoribbons (PNRs), Soleimanikahnoj
and Knezevic\citep{Soleimanikahnoj_et_al} and Forte\emph{ et al.}\citep{Forte_et_al}
studied the effect of vertical electric field on electronic and transport
properties of multilayer APNRs , while Yuan and Cheng\citep{Yuan_et_al}
investigated the influence of strain on the transport properties of
APNRs.

Guo \emph{et al}.\citep{Guo_et_al} studied the electronic structure
and geometries of bare and H-passivated APNRs using a first-principles
DFT based approach. Using the DFT\citep{Tran_et_al} Tran and Yang
studied the electronic structure and optical absorption of H-passivated
APNRs, and also reported scaling laws for their band gaps. Carvalho
and coworkers\citep{Carvalho_et_al} computed the formation energies
of APNRs, and also examined the edge-induced gap states in them, employing
DFT and analytical models. The influence of edge-passivation by chemical
groups such as H, F, Cl, O, S, Se, and OH, on the electronic properties
of APNRs was studied by Peng \emph{et al.}\citep{Peng_et_al}, using
the DFT. Maity and coworkers\citep{Maity_et_al} studied edge reconstruction
and Peierls transition in PNRs, using a DFT based approach. Wu \emph{et
al.}\citep{Wu_et_al} computed the electronic and transport properties
of H-passivated APNRs using a methodology combining DFT and nonequilibrium
Green\textquoteright s functions (NEGF.).\textcolor{teal}{{} }The electronic
structure and the Seeback coefficients of H-passivated PNRs, with
possible thermo-electric applications, were studied by Zhang \emph{et
al.}\citep{Zhang_et_al} also using the first-principles DFT. Hu,
Lin, and Yang\citep{Hu_et_al} studied the edge reconstruction in
unpassivated PNRs, including a few thousands to ten thousand atoms
in their calculations, by means of Discontinuous Galerkin DFT (DGDFT)
methodology. The optical properties of relatively narrow APNRs were
studied by Nourbakhsh and Asgari\citep{Nourbakhsh_et_al} , by going
beyond the first-principles DFT approach, by including electron-correlation
and particle-hole effects within G\textsubscript{0}W\textsubscript{0}
and Bethe-Salpeter equation (BSE) methodology. including excitonic
effects. Shekarforoush, Shiri and Khoeini\citep{Shekarforoush_et_al}
computed the linear and non-linear optical properties of H-passivated
APNRs of moderate widths, using first-principles DFT. Kaur \emph{et
al.}\citep{Kaur_et_al} also employed the first-principles DFT to
study the electronic, structural, and mechanical properties of PNRs
of several allotropes of phosphorene\textcolor{teal}{.} Electronic
structure of pristine and APNRs passivated by H, O, and OH wwas studied
by Ding and coworkers\citep{Ding_et_al} using first-principles DFT.
Possibility of using bilayer PNRs as pressure sensors was explored
by Lv \emph{et al.}\citep{Lv_et_al} theoretically, using the first-principles
DFT. Li and coworkers\citep{Li_et_al} also studied the electronic
structure of both pristine and H-passivated PNRs of moderate widths,
using the first-principles DFT. Using a similar computational approach,
Gueorguiev and coworkers have studied other lower-dimensional systems,
such as bismuth sheets, and carbon chains.\citep{Gueorguiev_1,Gueorguiev_2}

In this work, we have studied both the pristine and the passivated
APNRs ranging from very narrow ribbons of width 0.32 nm to very broad
ones of width 3.85 nm, using the PBE functional. The widths of nanoribbons
can also be denoted using an integer parameter $N$ (see Fig. \ref{fig:geometry}(a)),
in terms of which the APNRs studied in this work are in the range
$3\leq N\leq24$. The novel aspects of this work are: (a) the maximum
width of ribbons studied in this work is more than the earlier first-princples
DFT based studies, such as H-passivated APNRs of maximum width 2.50
nm by Tran et al.\citep{Tran_et_al}, and pristine and passivated
APNRs by Peng et al.\citep{Peng_et_al}, on selective widths up to
3.50 nm, (b) narrow pristine APNRs of widths $N=3-5$ have been studied,
for which no earlier work has been reported, (c) the optical absorption
spectra of pristine and all edge-passivated APNRs, whereas the earlier
calculations employing the same approach exist only for H-passivated
ribbons,\citep{Tran_et_al} (d) the bands involved in the transitions
leading to a few important peaks in the absorption spectrum have been
identified, (e) for obtaining more accurate estimates of the band
gaps, we also performed the band structure calculations using the
HSE06 hybrid functional for ribbons of medium width, with several
passivating groups, whereas earlier calculations were performed only
for -H, and -OH passivated ribbons.\citep{Ding_et_al} Additionally,
we also analyzed the calculated data of formation energies and concluded
that: (i) the smaller ribbons are more stable than the wider APNRs,
(ii) F- and OH-saturated structures are comparatively much more stable,
as compared to the H-passivated ones. Furthermore, we also performed
a detailed analysis of the contributions of various atoms to the the
orbitals in the frontier regions of valence and conduction bands.
To the best of our knowledge, none of the previous works has presented
calculations on electronic and optical properties of APNRs over such
a large range of width, for a variety of passivating groups.

The remaining part of this paper is organized as follows. The theoretical
approach and technical details are described in the next section.
We discuss and present our results in the section \ref{sec:results}.
Finally, we present our conclusions in section \ref{sec:conclusions}.

\section{Theoretical approach and Computational details}

\label{sec:theory}

The nanoribbons investigated in this work were taken to be periodic
along the $y$-direction, with at least 15 $\text{Å}$ vacuum included
in the supercell along both the $x$- and $z$- directions to minimize
the spurious interactions. First, geometry optimization for each ribbon
was performed, followed by the calculations of quantities such as
formation energies, the band gaps, the band structure, and the optical
absorption spectra. The calculations were carried out by employing
the first principles \emph{ab initio} comprehensive density functional
theory (DFT),\citep{Kohn_et_al} as implemented in the computer program
Vienna Ab-initio Simulation Package (VASP).\citep{Kresse_et_al,Kresse_2_et_al}
For the purpose, we used projector-augmented wave (PAW) pseudo-potentials\citep{Blochl_et_al,Kresse_3_et_al}
and Perdew-Burke-Ernzerhof (PBE) exchange-correlation functional\citep{Perdew_et_al}
for geometry optimization.

The kinetic energy cut-off of 500 eV was used for the plain wave basis
set. For geometry optimization, a k-point grid of 1$\times$14$\times$1
were chosen in the reciprocal space by employing Monkhorst-Pack centered
at the $\Gamma$ point. During geometry optimization, the convergence
cutoff for the electronic energy was $10^{-5}$ eV, while that for
Hellmann-Feynman force was set to $10^{-2}$ eV/$\text{Å}$. For subsequent
total energy and density of states calculations, a tighter energy
cutoff of $10^{-6}$ eV, along with a denser k-mesh grid of 1$\times$45$\times$1,
were employed.\textcolor{red}{{} }For band structure calculations of
the optimized APNRs, 100 k-points were included for the path from
$\Gamma$ to Y. Both the PBE and Heyd-Scuseria-Ernzerhof (HSE06)\citep{Heyd_et_al}
hybrid functionals ware employed for self-consistent electronic structure
calculations.

\section{Results and Discussion}

\label{sec:results}

In this section we present and discuss the results of our calculations
on the geometry, stability, electronic structure, and optical properties
of APNRs, with edges passivated by different functional groups.

\subsection{Geometry}

\begin{figure}
\begin{centering}
\subfloat[Edge-passivated APNR.]{\begin{centering}
\includegraphics[scale=0.3]{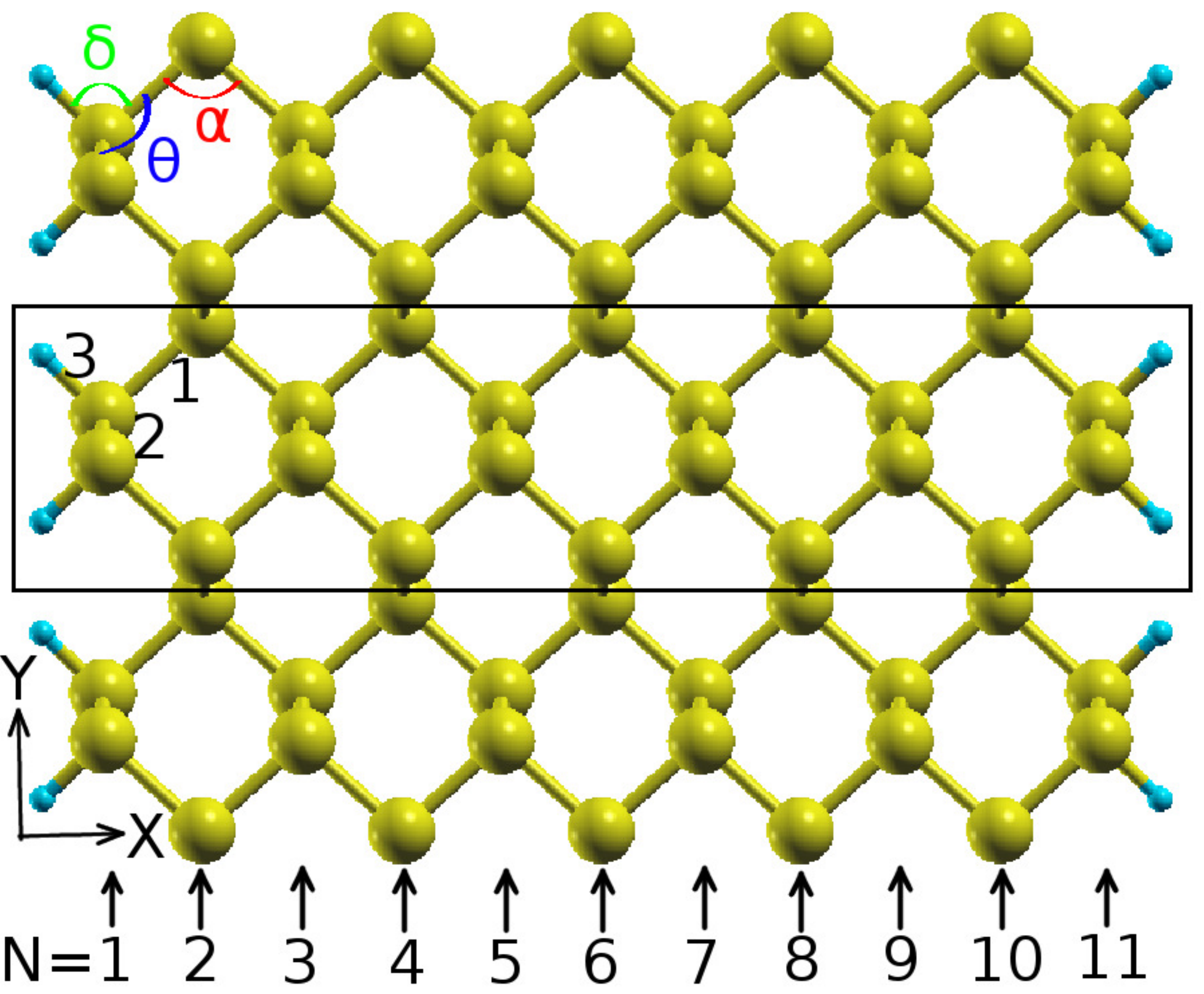}
\par\end{centering}
}~~\includegraphics[scale=0.15]{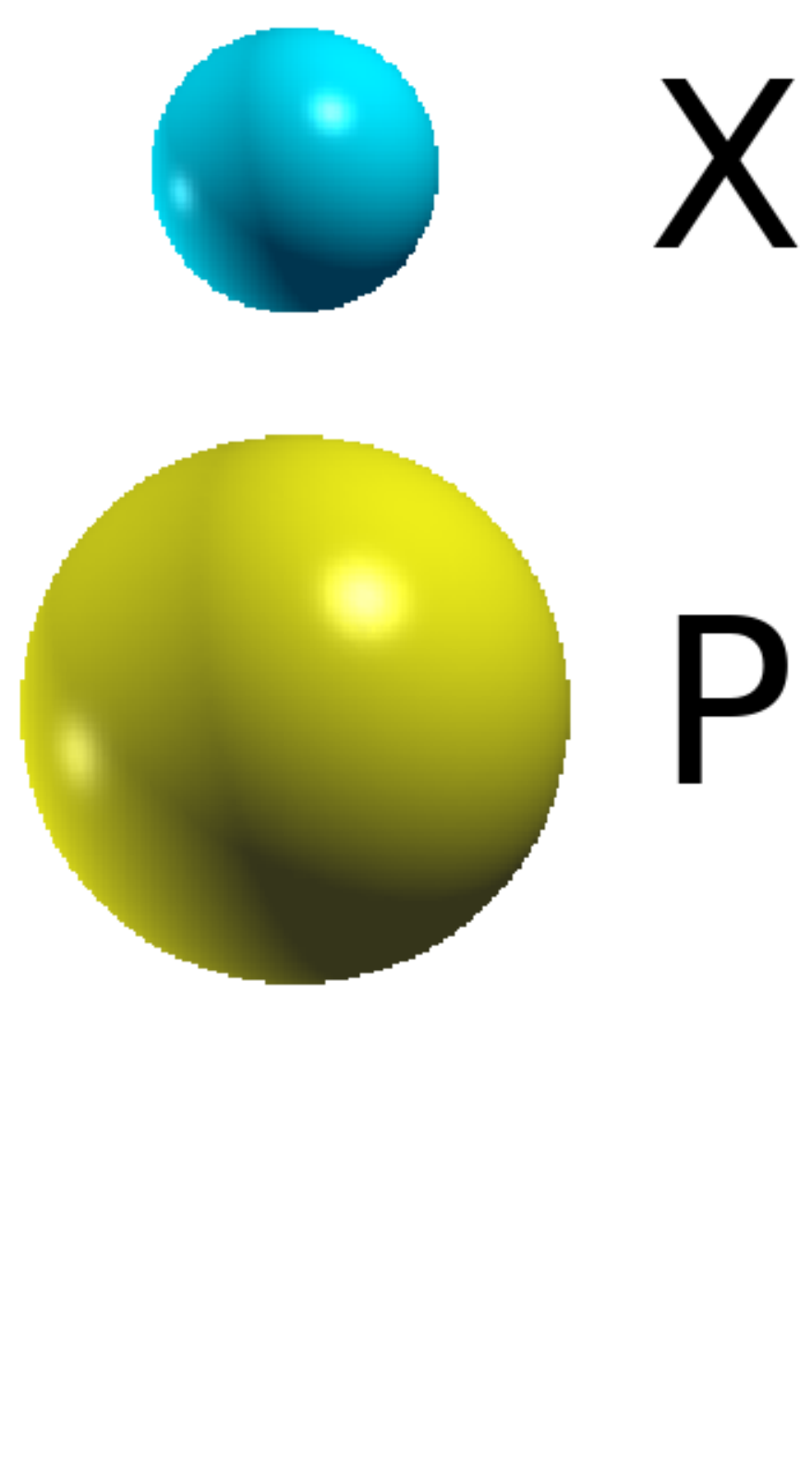}
\par\end{centering}
\begin{centering}
\subfloat[Pristine APNR]{\begin{centering}
\includegraphics[scale=0.27]{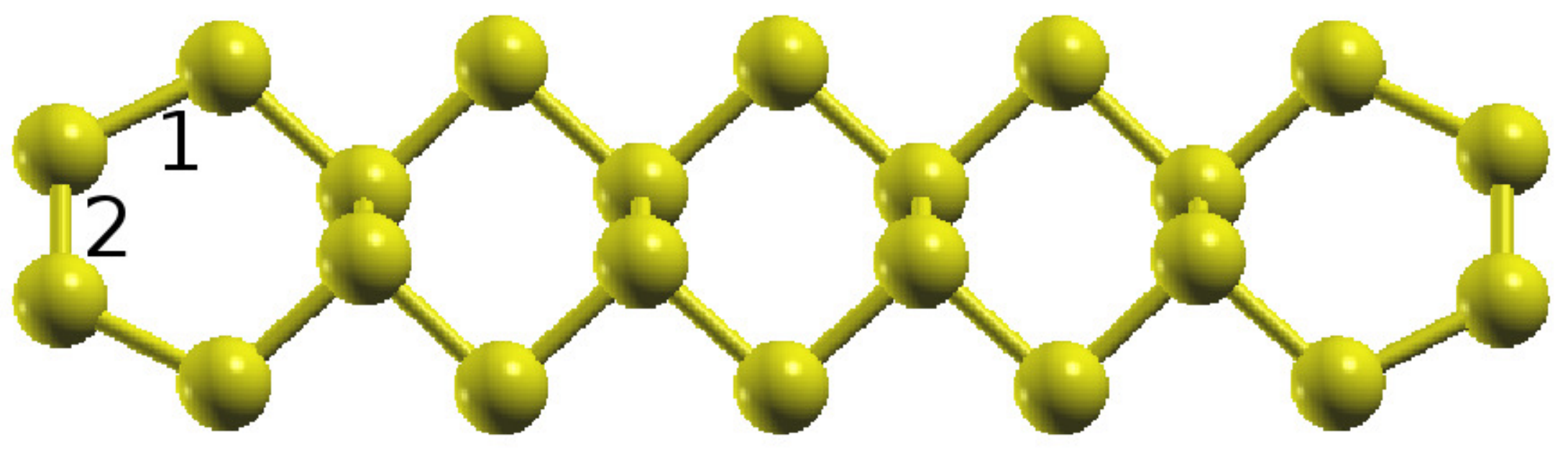}
\par\end{centering}
}
\par\end{centering}
\caption{Relaxed geometry of 11-APNRs with and without edge passivation, where
X denotes the passivating atom/group. 1, 2, and 3 indicate the bonds
corresponding to the bond lengths l\protect\textsubscript{1}, l\protect\textsubscript{2},
and l\protect\textsubscript{3} respectively, whose values are presented
in Table \ref{tab:geometry}.\label{fig:geometry}}
\end{figure}

Henceforth, we denote a given ribbon by its width $N$ (see Fig. \ref{fig:geometry}(a)),
thus an APNR of width $N$ will be denoted as $N$-APNR. Additionally,
edge bond angles and bond lengths relevant for both passivated and
pristine nanoribbons are also defined in Fig. \ref{fig:geometry}.
We performed geometry optimization for $N$-APNRs, with $3\leq N\leq24$,
for all functional groups except OH and F, for which we performed
calculations only for $N=11$. Optimized geometries for all functional
groups as well for the pristine case are presented in Table \ref{tab:geometry},
for $N=11$. Using the same methodology we performed geometry optimization
on phosphorene monolayer for reference, and the corresponding relaxed
geometry parameters are also presented in Table \ref{tab:geometry}.
We note that our optimized geometries both for APNRs and phosphorene
monolayer are in excellent agreement with results reported by Peng
et \emph{al.}\citep{Peng_et_al}, also based upon first-principles
DFT.

For the pristine APNRs, the P-P bond length, l\textsubscript{2} has
been reduced significantly as compared to the other structures in
order to stabilize the dangling bonds at the edges, and, because of
that, the bond angles $\alpha$ and $\theta$ have also increased
considerably. Generally, one observes the tendency that larger the
edge passivating atom/group, the longer the corresponding bond length.
In the present case also the tendency holds true in that when selenium
(Se) passivates the edge, l\textsubscript{3} is the longest as compared
to its values for other edge atoms. As far as interior bond lengths
and angles are concerned, they differ from their edge counterparts
by small amounts.

\begin{table}
\caption{The bond lengths (l\protect\textsubscript{1}, l\protect\textsubscript{2},
l\protect\textsubscript{3}) and the bond angles ($\alpha$, $\theta$,
$\delta$) at the edges of the APNRs are shown in the Fig.\ref{fig:geometry}
where the bond lengths are denoted according to their subscript 1,
2 and 3.\label{tab:geometry}}

\centering{}%
\begin{tabular}{ccccccccccccc}
\hline 
Structure &  & l\textsubscript{1} &  & l\textsubscript{2} &  & l\textsubscript{3} &  & $\alpha$ &  & $\theta$ &  & $\delta$\tabularnewline
 &  & ($\text{Å}$) &  &  ($\text{Å}$) &  & ($\text{Å}$) &  & ($^{\circ}$) &  & ($^{\circ}$) &  & ($^{\circ}$)\tabularnewline
\hline 
\hline 
Mono-layer &  & 2.22 &  & 2.26 &  & NA &  & 95.9 &  & 104.1 &  & NA\tabularnewline
Pristine &  & 2.23 &  & 2.07 &  & NA &  & 110.8 &  & 119.4 &  & NA\tabularnewline
H &  & 2.22 &  & 2.25 &  & 1.44 &  & 95.9 &  & 103.3 &  & 93.0\tabularnewline
OH &  & 2.25 &  & 2.26 &  & 1.64 &  & 96.9 &  & 99.9 &  & 103.2\tabularnewline
F &  & 2.23 &  & 2.25 &  & 1.64 &  & 94.7 &  & 97.8 &  & 98.6\tabularnewline
Cl &  & 2.24 &  & 2.26 &  & 2.07 &  & 92.1 &  & 96.1 &  & 101.7\tabularnewline
S &  & 2.23 &  & 2.25 &  & 2.12 &  & 97.1 &  & 104.9 &  & 99.5\tabularnewline
Se &  & 2.23 &  & 2.24 &  & 2.28 &  & 95.3 &  & 102.9 &  & 100.0\tabularnewline
\hline 
\end{tabular}
\end{table}

\subsection{Formation energy and relative stability}

\begin{figure}
\begin{centering}
\subfloat[\label{fig:form_energy-1}]{\begin{centering}
\includegraphics[scale=0.36]{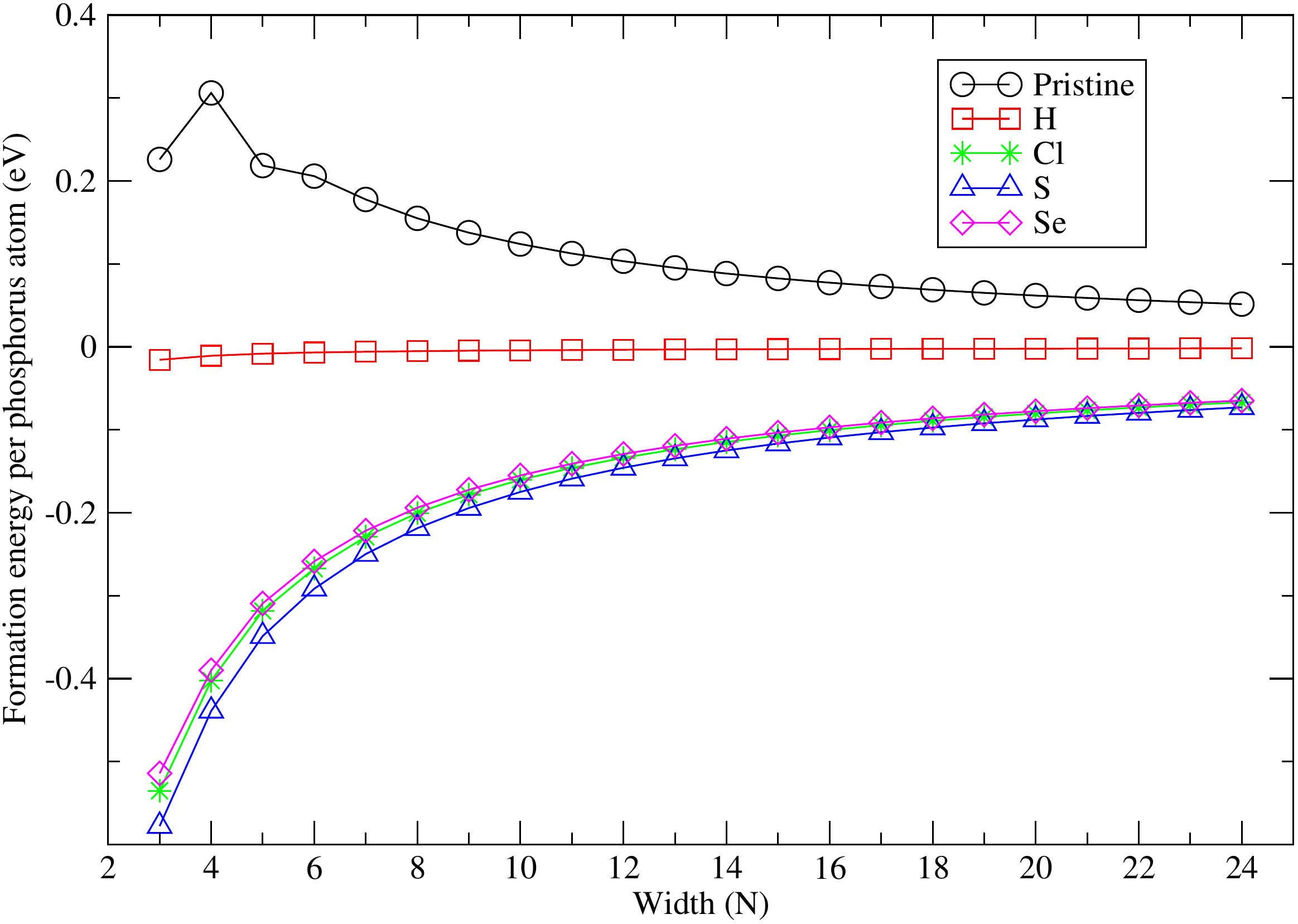}
\par\end{centering}
}\subfloat[\label{fig:form_energy}]{\begin{centering}
\includegraphics[scale=0.25]{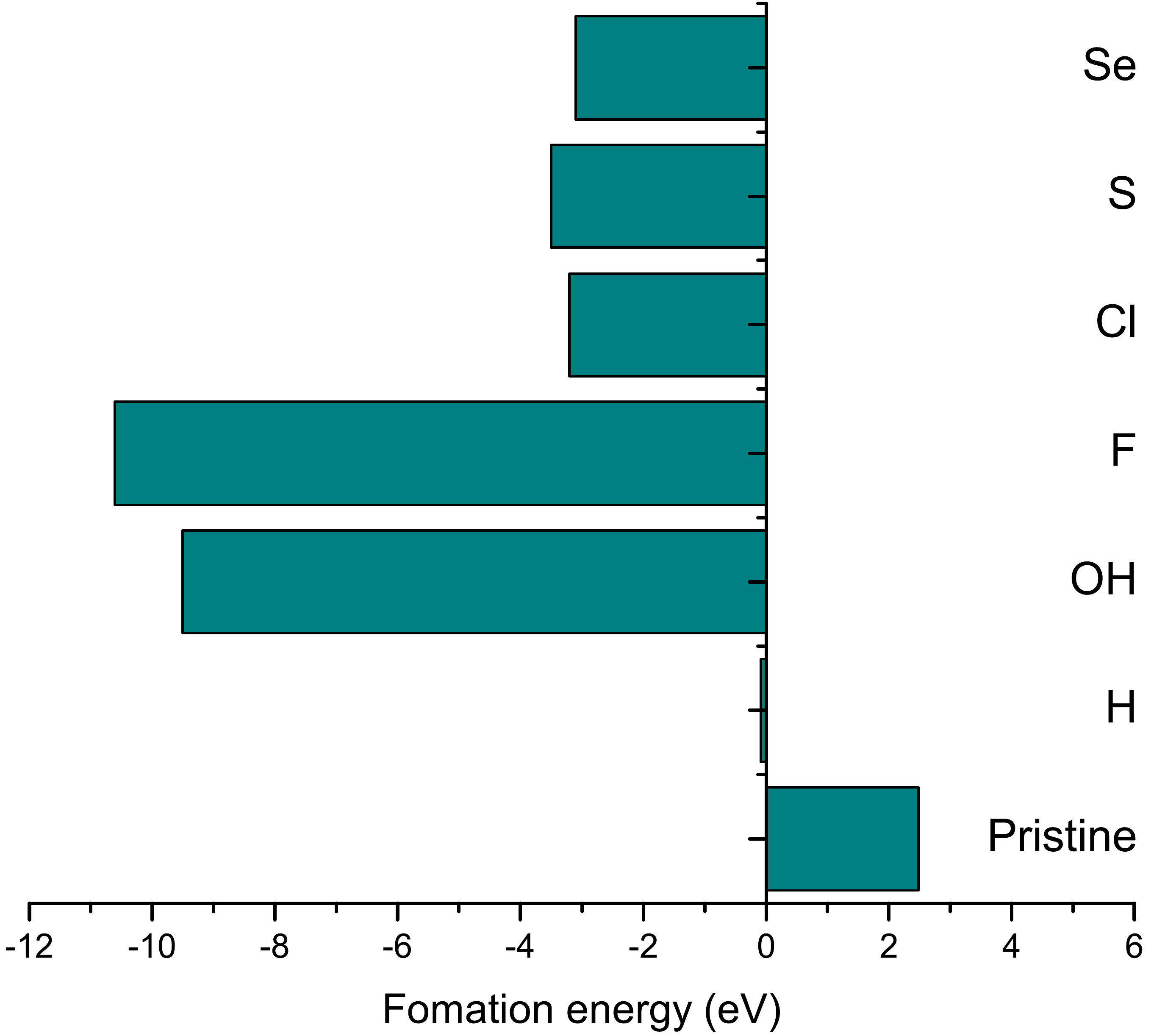}
\par\end{centering}
}
\par\end{centering}
\caption{Formation energies of APNRs calculated using the PBE functional: (a)
width dependence of formation energy per P-atom for pristine and various
edge-saturated cases, and (b) formation energies of 11-APNRs, for
various edge saturations.}
\end{figure}

To quantify the energetic stability of an APNR, we define its formation
energy ($E_{form}$) as follows

\begin{equation}
E_{form}=E_{total}-N_{p}E_{p}-N_{e}E_{e},\label{eq:1}
\end{equation}

where $E_{form}$ and $E_{total}$, respectively, are the formation
energy and total energy of a pure or edge-passivated APNR within a
super-cell containing $N_{p}$ phosphorus atoms, and $N_{e}$ passivation
atoms/groups at the edge. Furthermore, $E_{p}$ is the energy per
phosphorus atom of an infinite phosphorene sheet, and $E_{e}$ is
the energy of a edge-passivating foreign atom which we calculated
as $E_{e}=E_{e_{2}}/2$ . For the case of edge passivation by OH group,
we used $E_{e}=(E_{\mbox{O}_{2}}+E_{\mbox{H}_{2}})/2$. Thus, the
formation energy measures energetic stability of an edge-passivated
APNR, with respect to an infinite phosphorene sheet. Negative value
of $E_{form}$ clearly implies that the formation of a given APNR
from phosphorene sheet is energetically possible. In our calculations,
$N_{e}$=4 for all APNRs considered in this work (see Fig. \ref{fig:geometry}),
while $N_{p}$ depends on the width of the nanoribbon. For example,
for 11-APNR, $N_{p}$=22. Energies of all the molecules considered
as functional group are presented in the Table S1 of the supporting
information. In the pristine case, the third term on the right hand
side of Eq. \ref{eq:1} is absent. In Fig. \ref{fig:form_energy-1},
we present the plot of the width dependence of the formation energy
per phophorus (P) atom ($E_{form}/N_{p})$ for all the pristine and
edge-saturated APNR structures considered in this work, while the
formation energies of pristine as well as edge-passivated 11-APNRs
are plotted in Fig. \ref{fig:form_energy}. The following trends emerge
from these figures: (a) with positive values of $E_{form}$, pristine
APNRs are energetically unstable, however, their values of $E_{form}/N_{p}$
are decreasing with the width, i.e., they are reaching the limit of
infinite sheet, (b) for H-passivated ribbons, the formation energies
are negative, but much smaller than those for APNRs passivated by
other groups, (c) for all edge-passivated APNRs, the formation energy
per P-atom is gradulally increasing with the ribbon-width, and saturating
for larger widths, implying that the narrower saturated ribbons are
more stable than the broader ones. For the specific case of 11-APNR
(see Fig. \ref{fig:form_energy}), we note that the F-passivated APNR
is most stable, closely followed by the OH passivated ribbon. These
results of ours are consistent with those reported by Ding\emph{ et
al}.\citep{Ding_et_al}.

\subsection{Band gaps}

In this section we discuss the band gaps of APNRs as functions of
their width, and edge-passivating groups.

\begin{figure}
\begin{centering}
\includegraphics[scale=0.35]{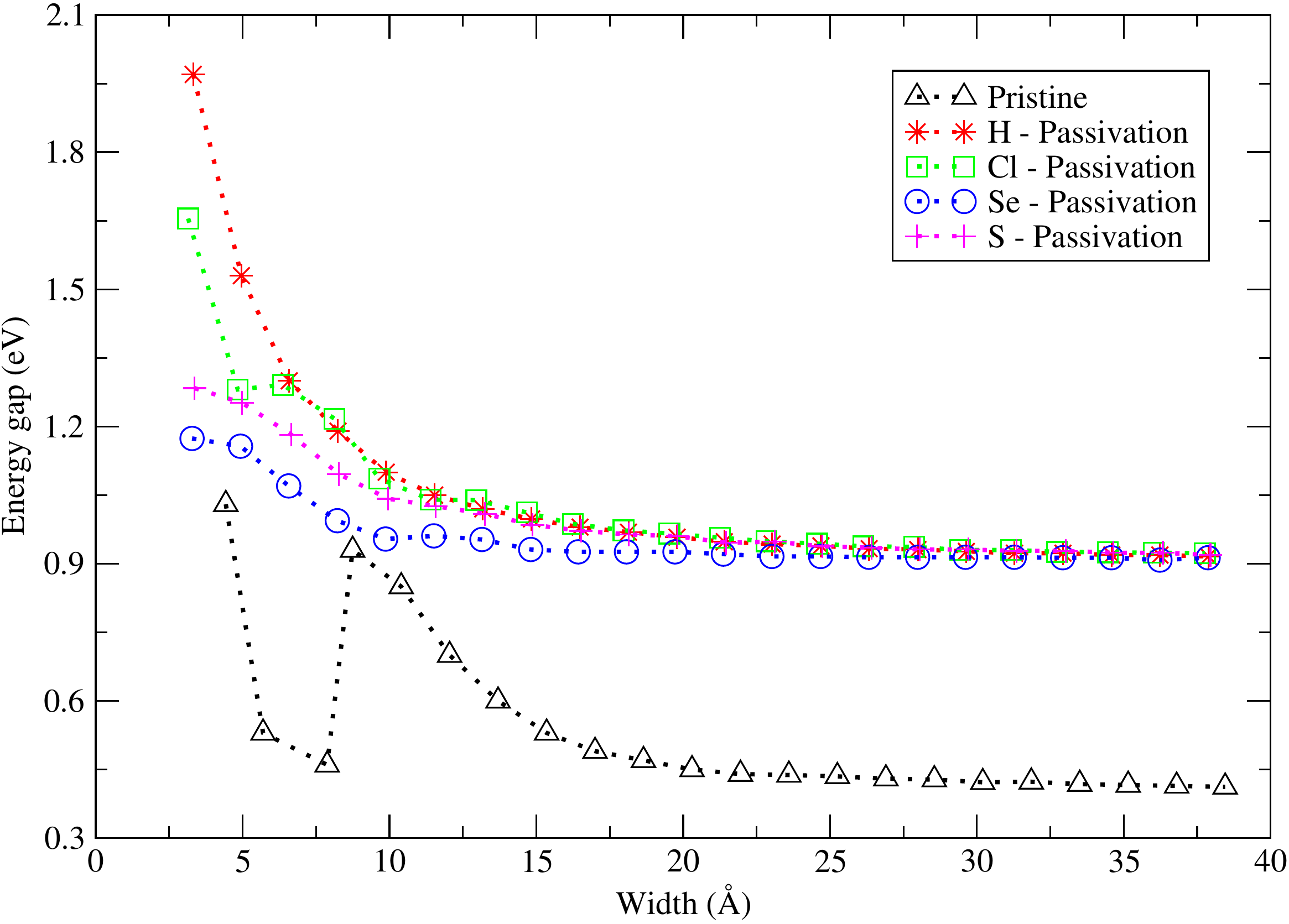}
\par\end{centering}
\caption{Variation of Band-gap of APNRs (up to 3.85 nm) with respect to the
ribbon width.\label{fig:Band-gap}}
\end{figure}

\begin{table}
\caption{Band gaps (in eV) of $N$-APNRs ($N=3-9$), for various passivating
atoms, calculated using the PBE functional.\label{tab:Band-gaps-passivation}}

\centering{}%
\begin{tabular}{|c||c|c|c|c|c|c|c|c|}
\hline 
\multicolumn{2}{|c|}{Passivating} & \multicolumn{7}{c|}{$N$}\tabularnewline
\cline{3-9} \cline{4-9} \cline{5-9} \cline{6-9} \cline{7-9} \cline{8-9} \cline{9-9} 
\multicolumn{2}{|c|}{Atom} & 3 & 4 & 5 & 6 & 7 & 8 & 9\tabularnewline
\hline 
\multicolumn{2}{|c|}{H} & 1.97 & 1.53 & 1.30 & 1.19 & 1.10 & 1.05 & 1.02\tabularnewline
\hline 
\multicolumn{2}{|c|}{Cl} & 1.66 & 1.28 & 1.29 & 1.22 & 1.09 & 1.04 & 1.04\tabularnewline
\hline 
\multicolumn{2}{|c|}{S} & 1.28 & 1.25 & 1.18 & 1.10 & 1.04 & 1.03 & 1.01\tabularnewline
\hline 
\multicolumn{2}{|c|}{Se} & 1.17 & 1.16 & 1.07 & 0.99 & 0.95 & 0.96 & 0.95\tabularnewline
\hline 
\end{tabular}
\end{table}

In Fig.\ref{fig:Band-gap}, we present the band gaps of APNRs as functions
of their widths, and our results are in good agreement with those
reported in earlier studies.\citep{Peng_et_al,Tran_et_al} An examination
of the figure reveals the following general trends: (a) with the increasing
width, the band gaps show a decreasing trend approaching saturation
around 2.5 nm, (b) for a given width, the band gap of a pristine nanoribbon
is smaller than that of a passivated one, and the difference grows
with the increasing width, before eventually saturating, (c) for a
given width, H-passivated ribbons have the largest band gaps, while
those passivated with Se have the smallest. But for the pristine ribbons
a peculiar behavior is observed in that the band gaps do not exhibit
a monotonic decrease with increasing width, for narrower ribbons.
At first, the band-gap decreases and reaches close to the final saturation
value, then suddenly increases again, eventually exhibiting a normal
decreasing trend with respect to the width. and follows the regular
trend. This is mainly due to the huge distortion in the smaller structures
to stabilize the dangling bonds which make changes in the symmetry
of wave function of the edge states. To the best of our knowledge,
so far there is no literature on on the electronic structure of pristine
APNRs, with widths in the range $N=3-5$.

It is also instructive to compare our obtained band gaps with that
of infinite phosphorene monolayer. Using the PBE functional, and the
geometry parameters listed in Table \ref{tab:geometry}, for the monolayer
phosphorene we obtained the band gap to be 0.91 eV. This compares
well with our saturated values of 0.92 eV for the H-passivated APNRs,
and 0.91 for the selenium passivated ribbons. However, it is significantly
larger than the saturated band gap value of 0.40 eV obtained for the
pristine APNRs, thereby implying that pristine APNRs, due to their
distorted edges, do not correctly evolve into monolayer phosphorene,
with the increasing width. To examine the effect of spin-orbit coupling,
we carried out calculations for hydrogen saturated 5- and 11-APNRs,
but no significant changes were observed.

\begin{table}
\caption{Comparison of the band gaps of pristine and edge-saturated 11-APNRs
computed using PBE and HSE06 hybrid functionals\label{tab:Band-gap-comparison}}

\centering{}%
\begin{tabular}{ccccc}
\hline 
Structure & \hspace{0.5cm} & Band-gap (eV) & \hspace{0.5cm} & Band-gap\tabularnewline
\cline{3-3} 
 &  & %
\begin{tabular}{ccc}
PBE & \hspace{0.2cm} & HSE06\tabularnewline
\end{tabular} &  & type\tabularnewline
\hline 
\hline 
Pristine &  & %
\begin{tabular}{ccc}
0.49 & \hspace{0.5cm} & 1.05\tabularnewline
\end{tabular} &  & Indirect\tabularnewline
H &  & %
\begin{tabular}{ccc}
0.98 & \hspace{0.5cm} & 1.68\tabularnewline
\end{tabular} &  & direct\tabularnewline
OH &  & %
\begin{tabular}{ccc}
0.99 & \hspace{0.5cm} & 1.68\tabularnewline
\end{tabular} &  & direct\tabularnewline
F &  & %
\begin{tabular}{ccc}
0.96 & \hspace{0.5cm} & 1.65\tabularnewline
\end{tabular} &  & direct\tabularnewline
Cl &  & %
\begin{tabular}{ccc}
0.99 & \hspace{0.5cm} & 1.68\tabularnewline
\end{tabular} &  & direct\tabularnewline
S &  & %
\begin{tabular}{ccc}
0.97 & \hspace{0.5cm} & 1.67\tabularnewline
\end{tabular} &  & direct\tabularnewline
Se &  & %
\begin{tabular}{ccc}
0.92 & \hspace{0.5cm} & 1.64\tabularnewline
\end{tabular} &  & direct\tabularnewline
\hline 
\end{tabular}
\end{table}

It is well-known that DFT-PBE based approaches generally underestimate
the bang gaps significantly, therefore, one can wonder as to how reliable
are our PBE functional based calculations, both quantitatively, and
qualitatively. To verify that, we performed band gap calculations
on 11-APNRs for pristine as well as various edge-passivated configurations
using the HSE06 functional, and the results are presented in Table
\ref{tab:Band-gap-comparison}, along with the corresponding values
obtained using the PBE functional. HSE06 functional belongs to the
class of hybrid functionals,\citep{hse-1,hse-2} whose predicted band
gap values are normally fairly accurate, and compare well with the
experiments.\citep{Pela_HSE_comp_2015,Garza_HSE_comp_2016} We note
the following: (a) HSE06 band gap values for all the cases are significantly
higher as compared to the PBE values, and (b) the trends in the band
gap values ranging from pristine ribbons to the ones passivated by
Se atom are similar for both the HSE06 and PBE calculations. Therefore,
we conclude that although PBE functional significantly underestimates
the band gaps, however it reproduces the qualitative features of the
behavior of band gaps with respect to the passivation groups. Based
upon these results, and the data presented in Fig. \ref{fig:Band-gap}
and Table \ref{tab:Band-gaps-passivation}, we can say with certainty
that for the narrower APNRs, band gaps depend very sensitively on
the nature of edge passivation; those passivated by H atoms have the
largest gaps, while Se passivated ones have the smallest gaps. This
information can be used to tune the electronic and optical properties
of APNRs.

\subsection{Band structure and density of states (DOS)}

\begin{figure}
\begin{centering}
\includegraphics[scale=0.5]{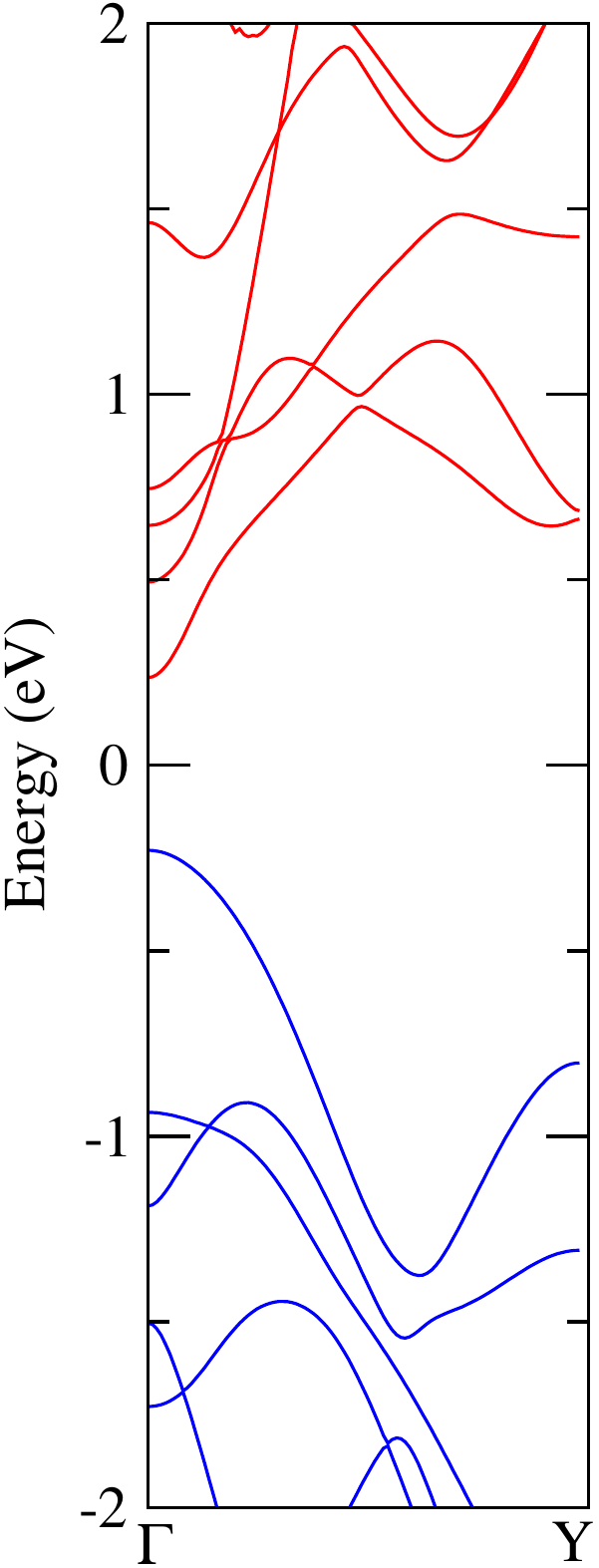}
\par\end{centering}
\caption{Band-structures of pristine 5-APNR\label{fig:Band-pristine-5-APNR}}
\end{figure}

\begin{figure}
\begin{centering}
\subfloat[Pristine]{\includegraphics[scale=0.4]{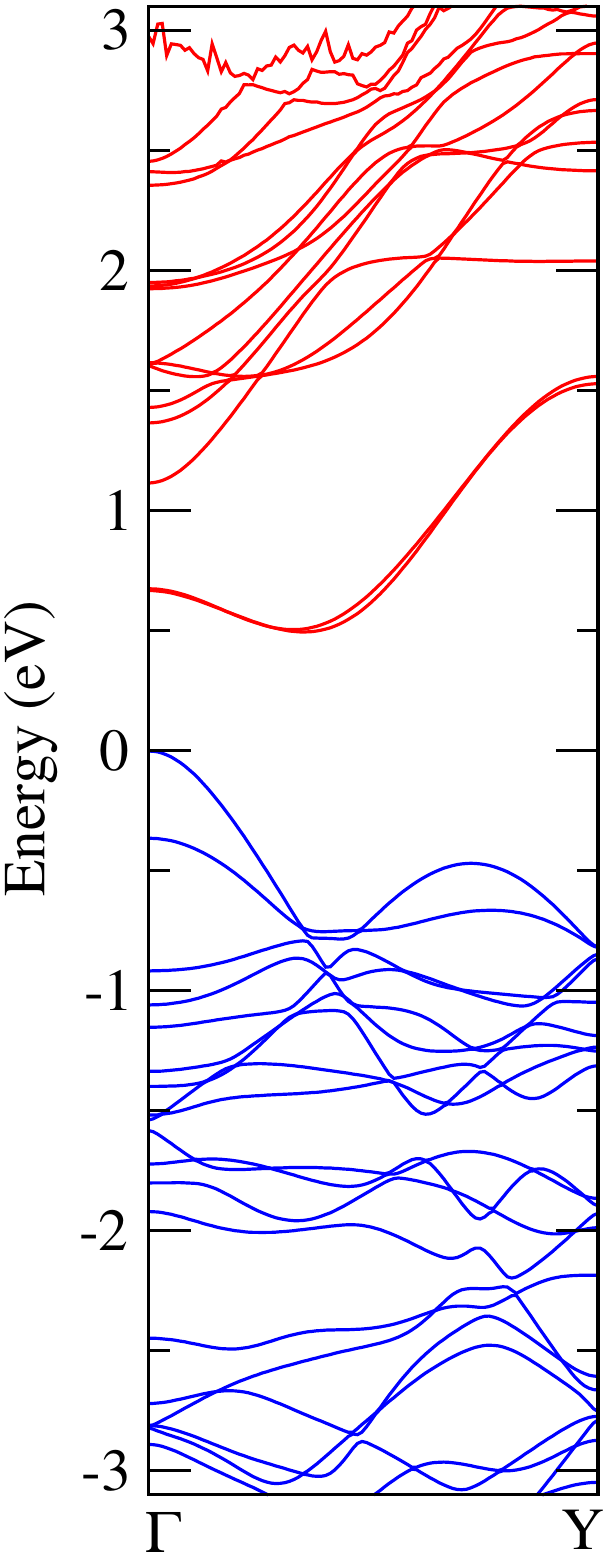}

}\subfloat[H]{\includegraphics[scale=0.4]{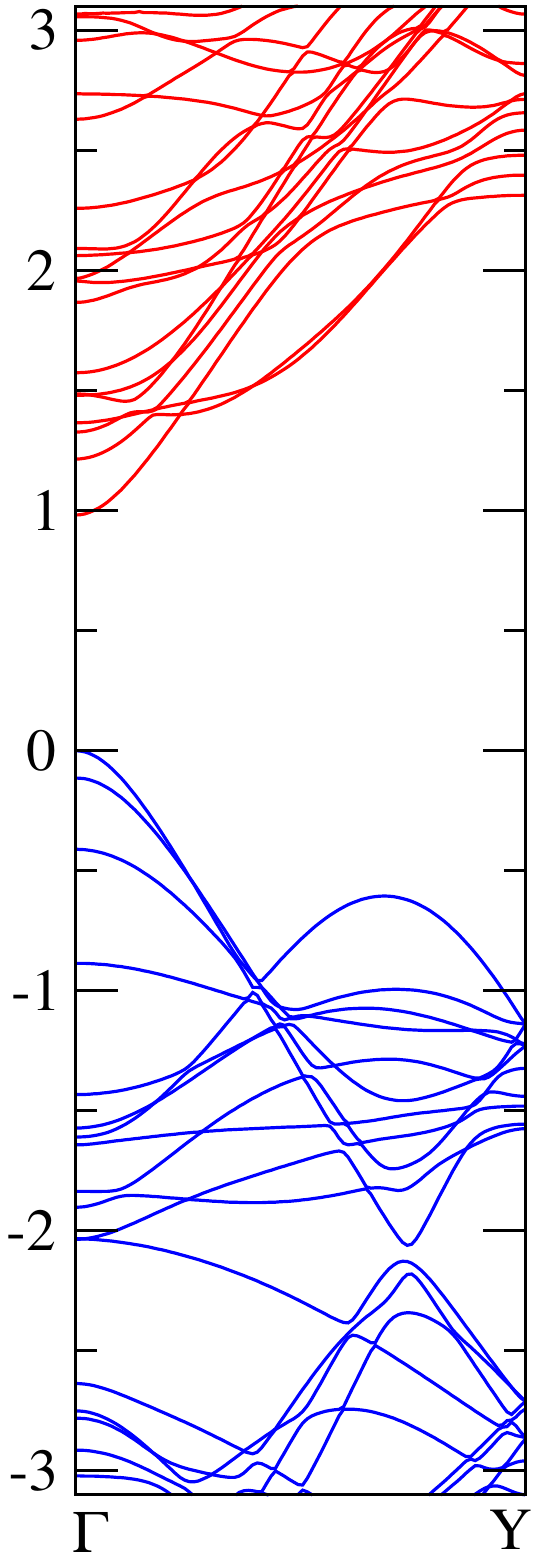}

}\subfloat[OH]{\includegraphics[scale=0.4]{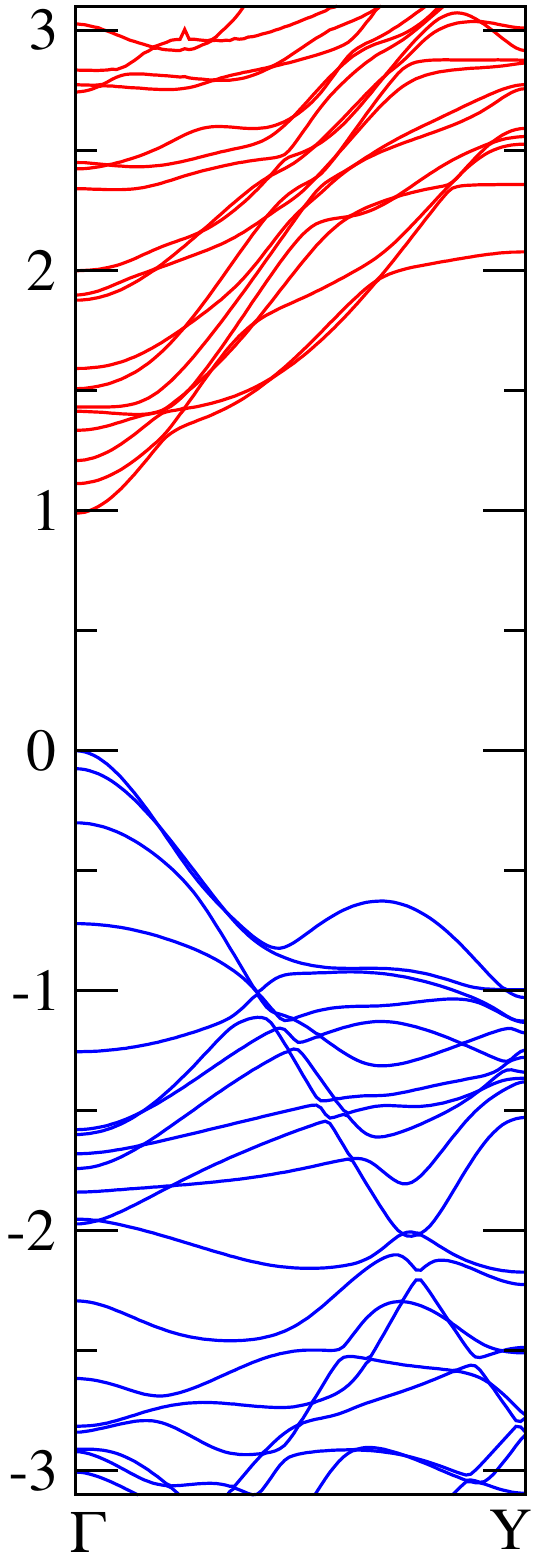}

}\subfloat[F]{\includegraphics[scale=0.4]{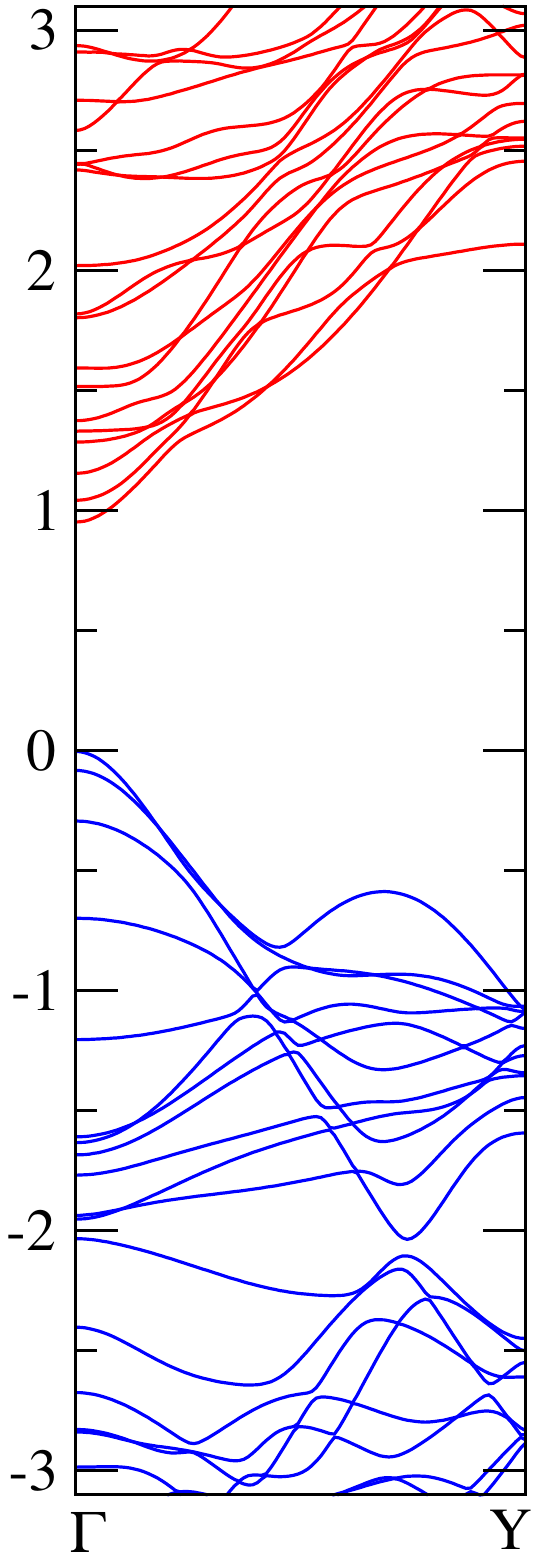}

}\subfloat[Cl]{\includegraphics[scale=0.4]{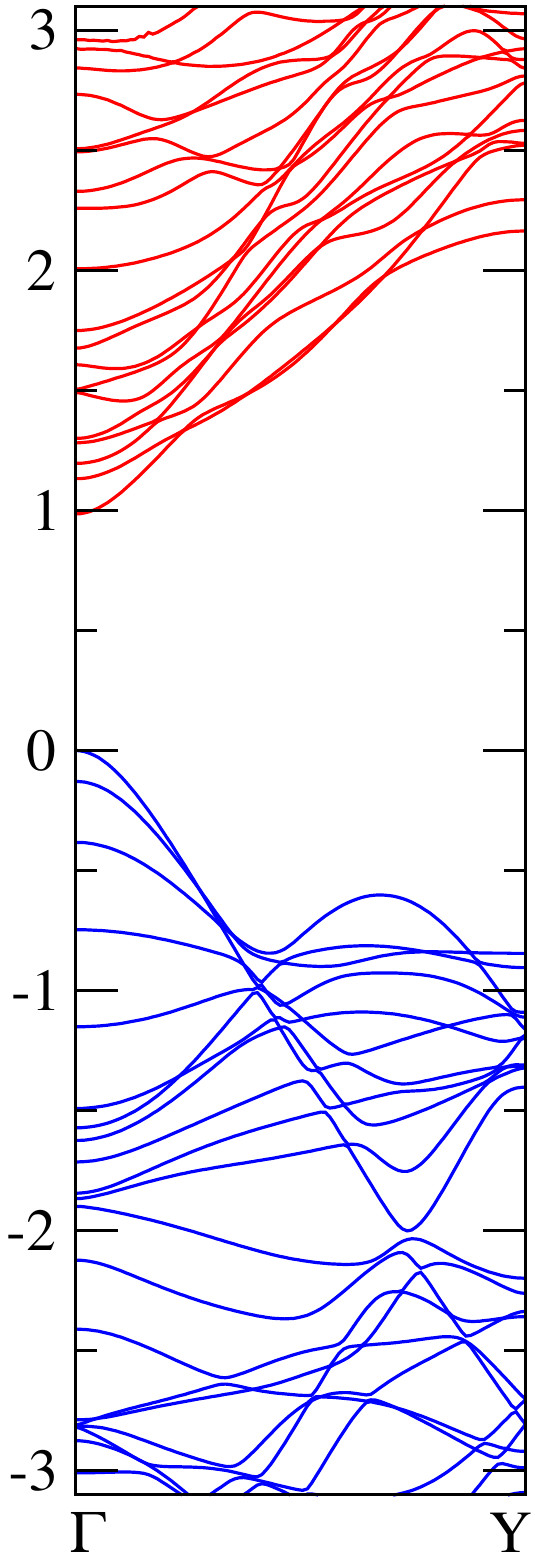}

}\subfloat[S]{\includegraphics[scale=0.4]{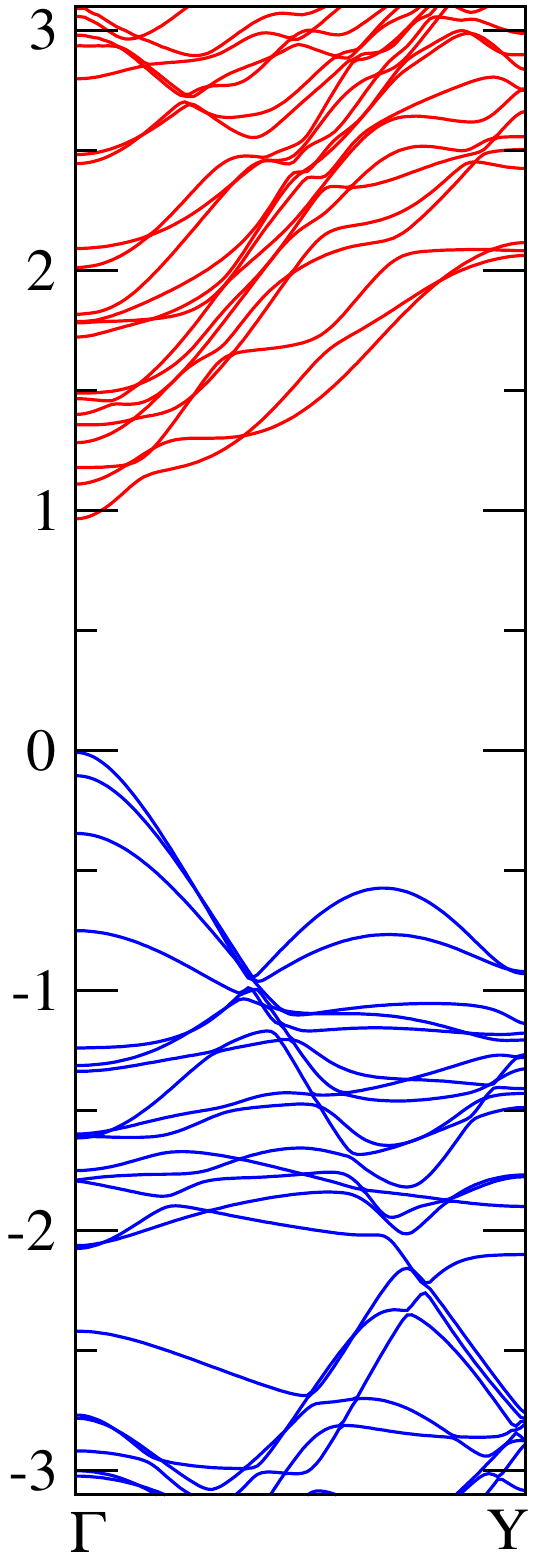}

}\subfloat[Se]{\includegraphics[scale=0.4]{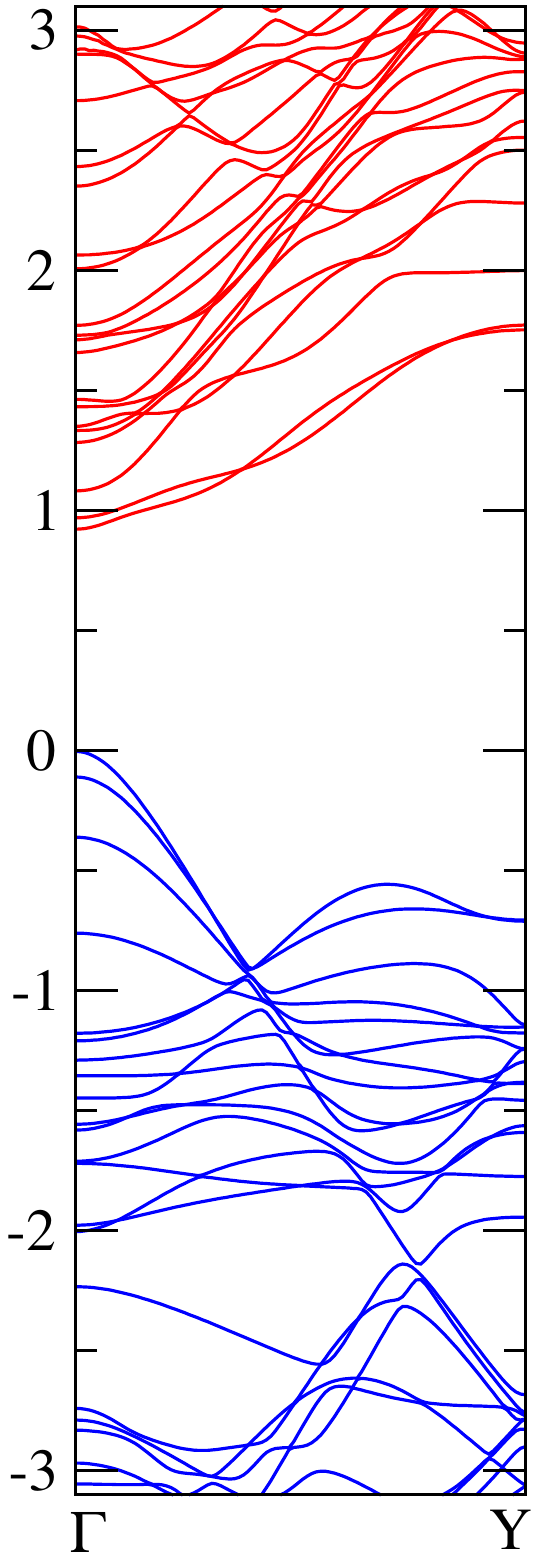}

}
\par\end{centering}
\caption{Band-structures of pristine and passivated 11-APNRs. The blue and
red curves are the states corresponding to the valence and the conduction
bands, respectively.\label{fig:Band-structures}}
\end{figure}

The computed band structures of pristine as well as passivated 11-APNR
are shown in the Fig. \ref{fig:Band-structures}. Our calculations
reveal that the band gaps of pristine APNRs of all but one width are
indirect, with the valence band maximum (VBM) at $\Gamma$ point,
and the conduction band minimum (CBM) between the $\Gamma$ and Y
points. The only exception to this is 5-APNR which, as shown in Fig.
\ref{fig:Band-pristine-5-APNR}, is a semiconductor with a direct
band gap of about 0.46 eV, at the $\Gamma$ point. To the best of
our knowledge, the band structure of pristine 5-APNR has not been
discussed in the literature earlier, and the reasons behind its direct
band gap could lie in its geometry, presented in Fig. S1 of the Supporting
Information. The bond lengths l\textsubscript{1} and l\textsubscript{2}
of 5-APNR are 2.30 $\text{Å}$ and 2.29 $\text{Å}$, respectively,
which are elongated with respect to the wider pristine structures.
As far as bond angles are concerned, the calculated value of $\alpha$
is 121.4$^{\text{°}}$, which is larger as compared to wider APNRs,
while the value of $\theta$ at 110$^{\circ}$ is smaller as compared
to the wider ribbons. 

The passivated nanoribbons, irrespective of their width, or the nature
of the passivating groups, turn out to be direct band gap semiconductors,
with the VBMs and CBMs located at the high symmetry $\Gamma$ point.
This is obvious from the Fig. \ref{fig:Band-structures}, which contains
the band structures of 11-APNRs passivated by H, OH, F, Cl, S, and
Se. These results of ours are in very good qualitative and quantitative
agreement with the calculations of Peng \emph{et al}.\citep{Peng_et_al}
and Tran and Yang\citep{Tran_et_al}. We performed calculations including
the spin-orbit coupling for hydrogen passivated 5- and 11-APNRs, and
observed no significant changes in the band structures.

For the edge-passivated APNRs, we find that both the VBM and the CBM
derive predominant contributions from p-type orbitals of P atoms located
both on the edges, as well as the interior of the ribbons. However,
for the Se-passivated APNRs, the lowest conduction band, including
the CBM, is mainly composed of the p-type orbitals located on the
Se atoms. As far as deeper valence band orbitals are concerned, they
derive dominant contribution from the s-type orbitals of the P atoms.
The contributions of passivating atoms to two lowest conduction bands
increase with the increasing atom sizes, being negligible for H-passivated
ribbons, and eventually reaching maximum values for Se-passivated
APNRs. A possible reason behind this behavior is that the electron
clouds of larger passivating atoms are extended further into the interior
of the nanoribbon, allowing its hybridization with the lower conduction
bands leading to their lowering, thus causing a reduction in the band
gaps. 

We have compared the band structure of H-passivated 11-APNR computed
using the HSE06 functional, with that obtained from the PBE functional,
in Fig. S4 of the Supporting Information. The HSE06 band structure
calculations were initiated using the PBE wave functions, and from
the figure it is obvious that the HSE bands close to the Fermi level
are similar to the PBE ones, except undergoing a rigid shift resulting
from the widening of the band gap. Similar trends were also observed
in the HSE band structures of APNRs passivated by other groups.

\begin{figure}
\begin{centering}
\subfloat[Pristine]{\includegraphics[scale=0.25]{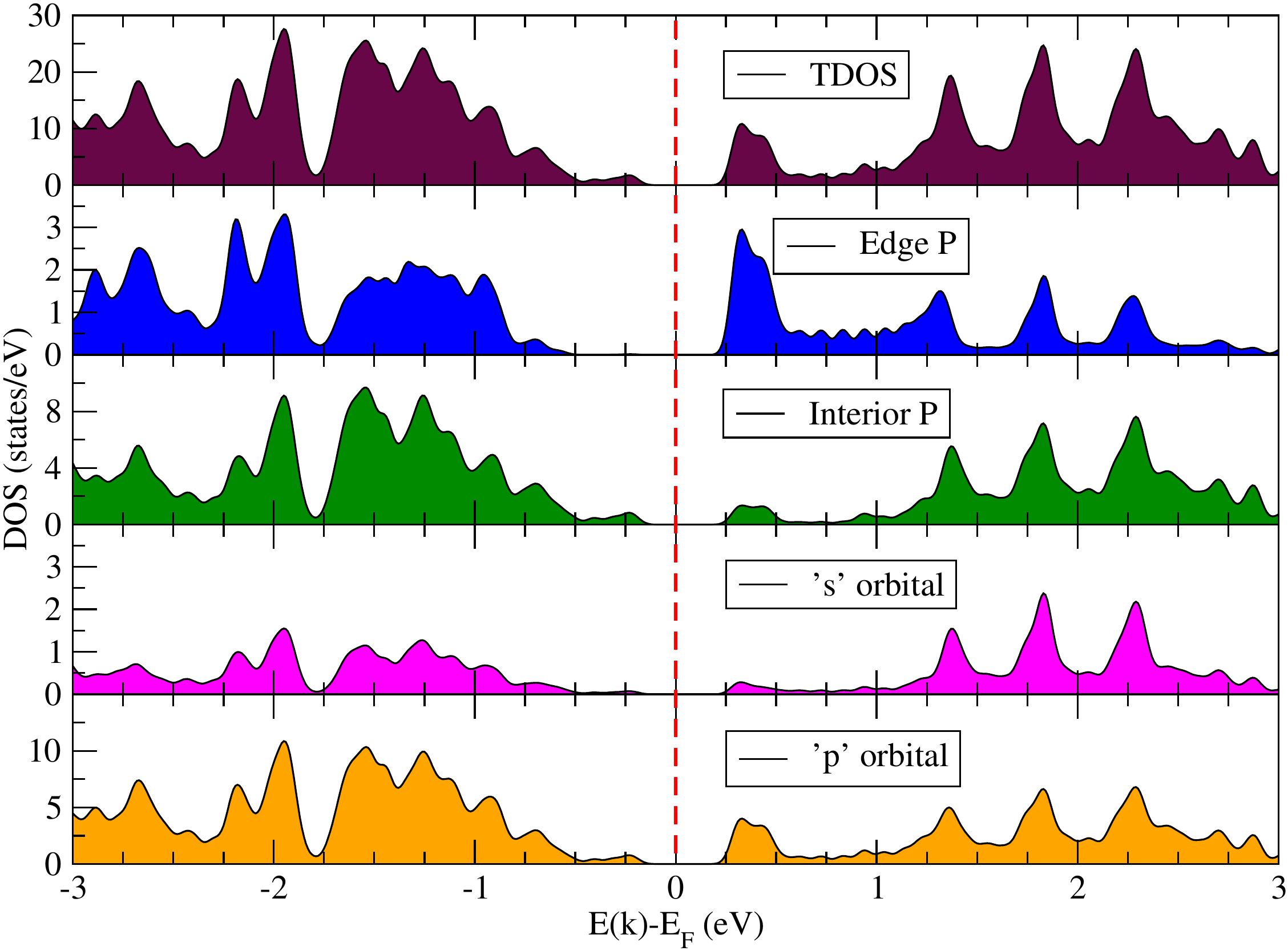}

}\subfloat[H]{\includegraphics[scale=0.25]{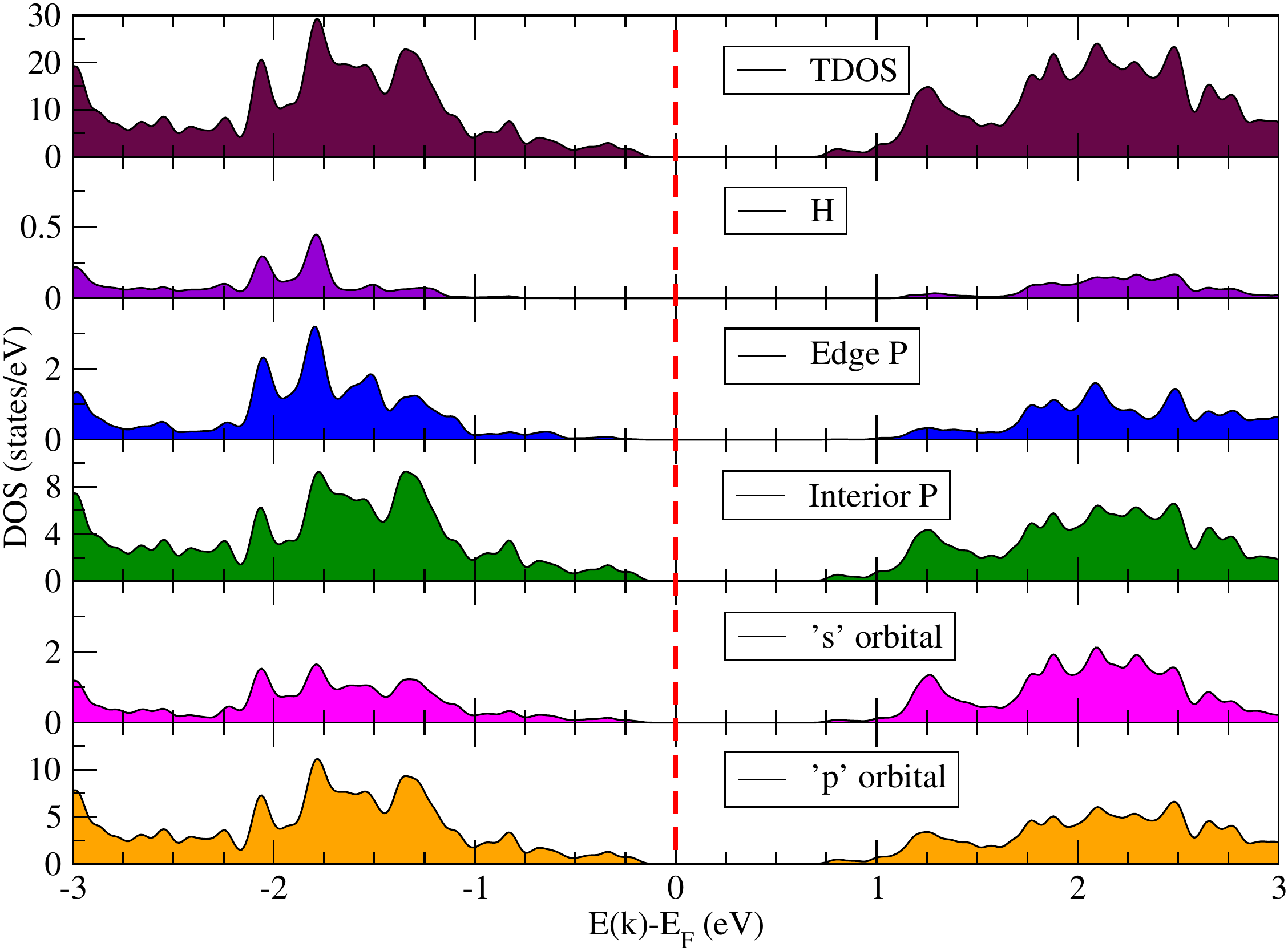}

}\subfloat[OH]{\includegraphics[scale=0.25]{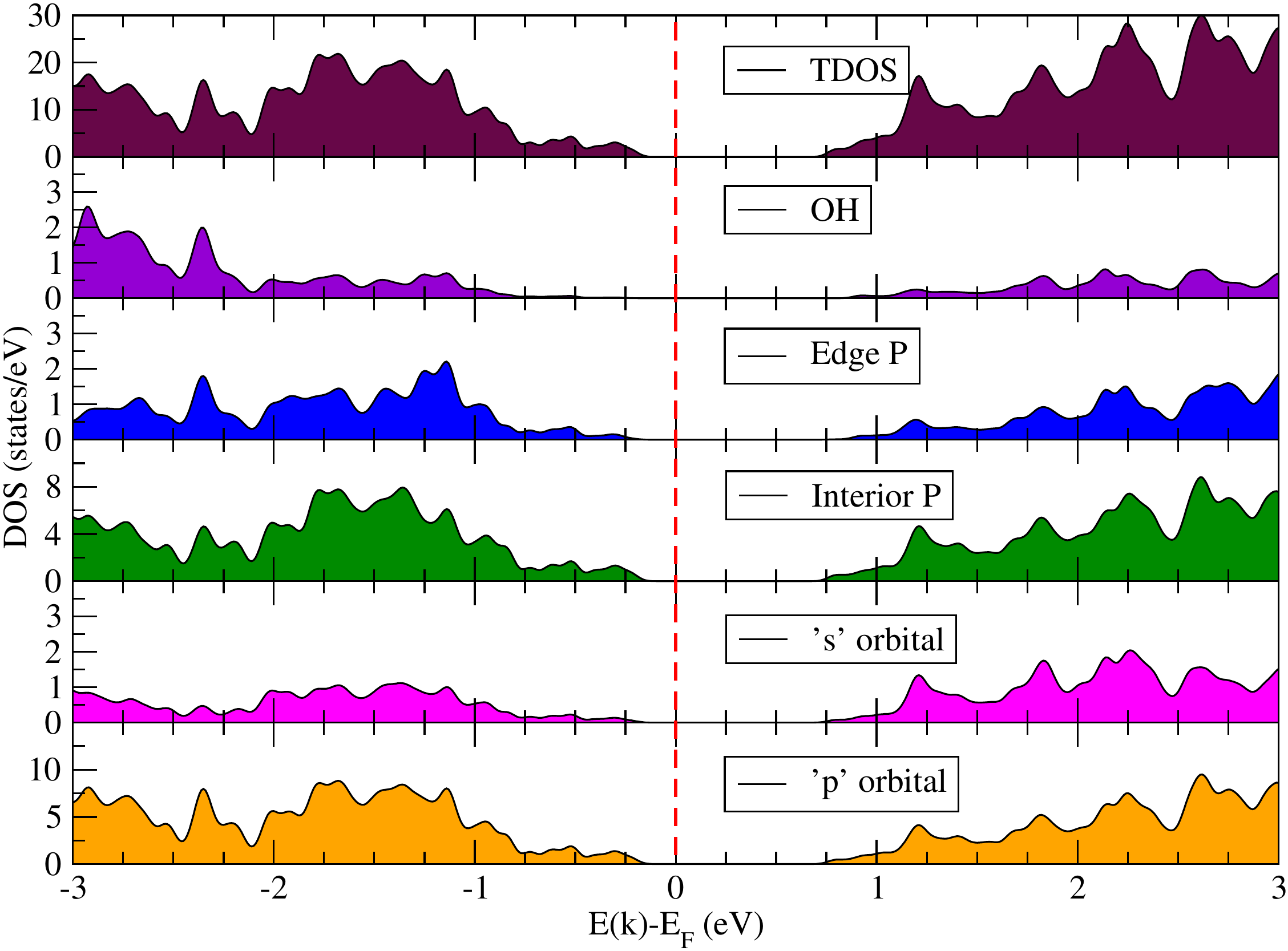}

}
\par\end{centering}
\begin{centering}
\subfloat[F]{\includegraphics[scale=0.25]{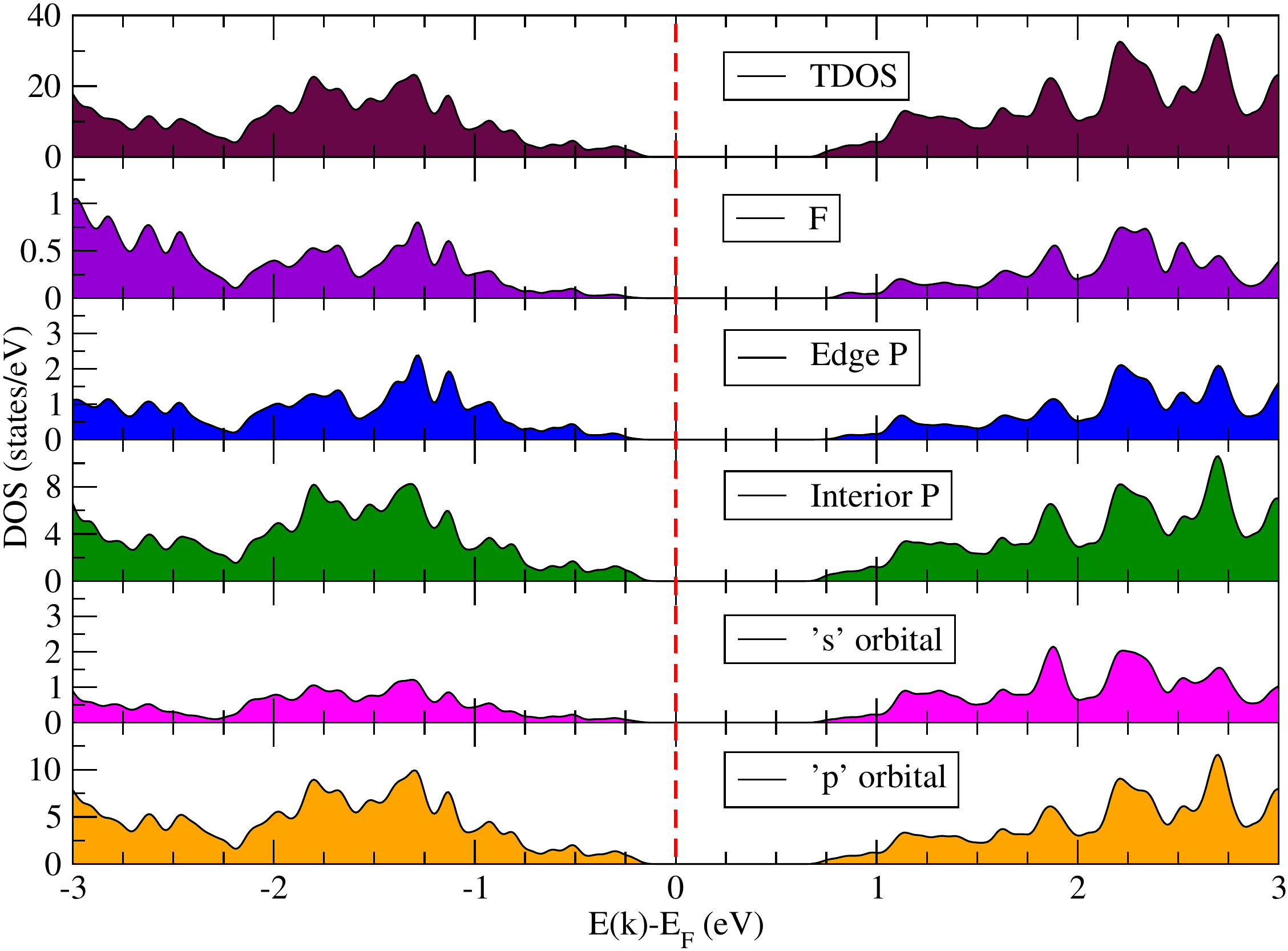}

}~~~~~~~~~\subfloat[Cl]{\includegraphics[scale=0.25]{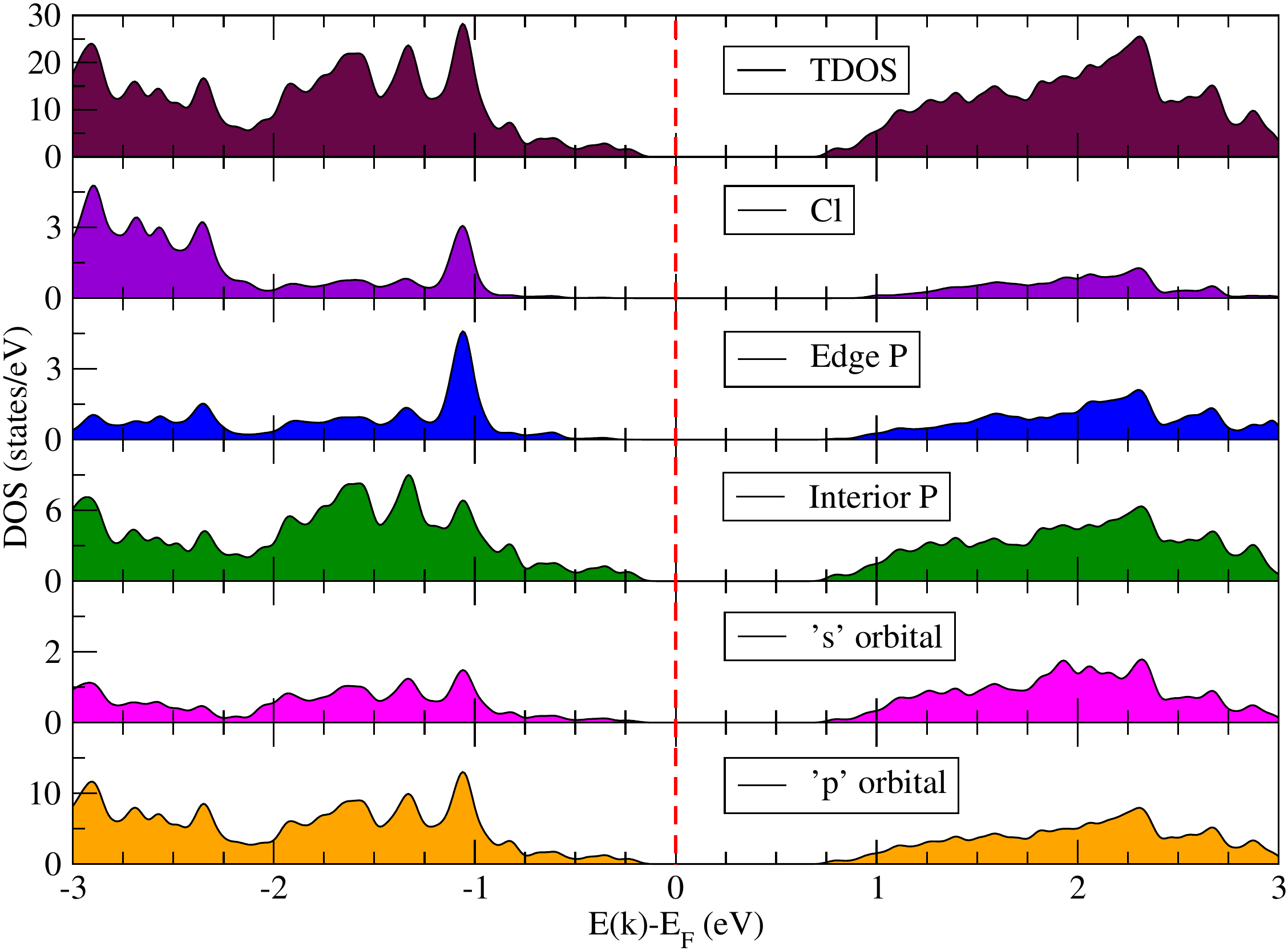}

}
\par\end{centering}
\begin{centering}
\subfloat[S]{\includegraphics[scale=0.25]{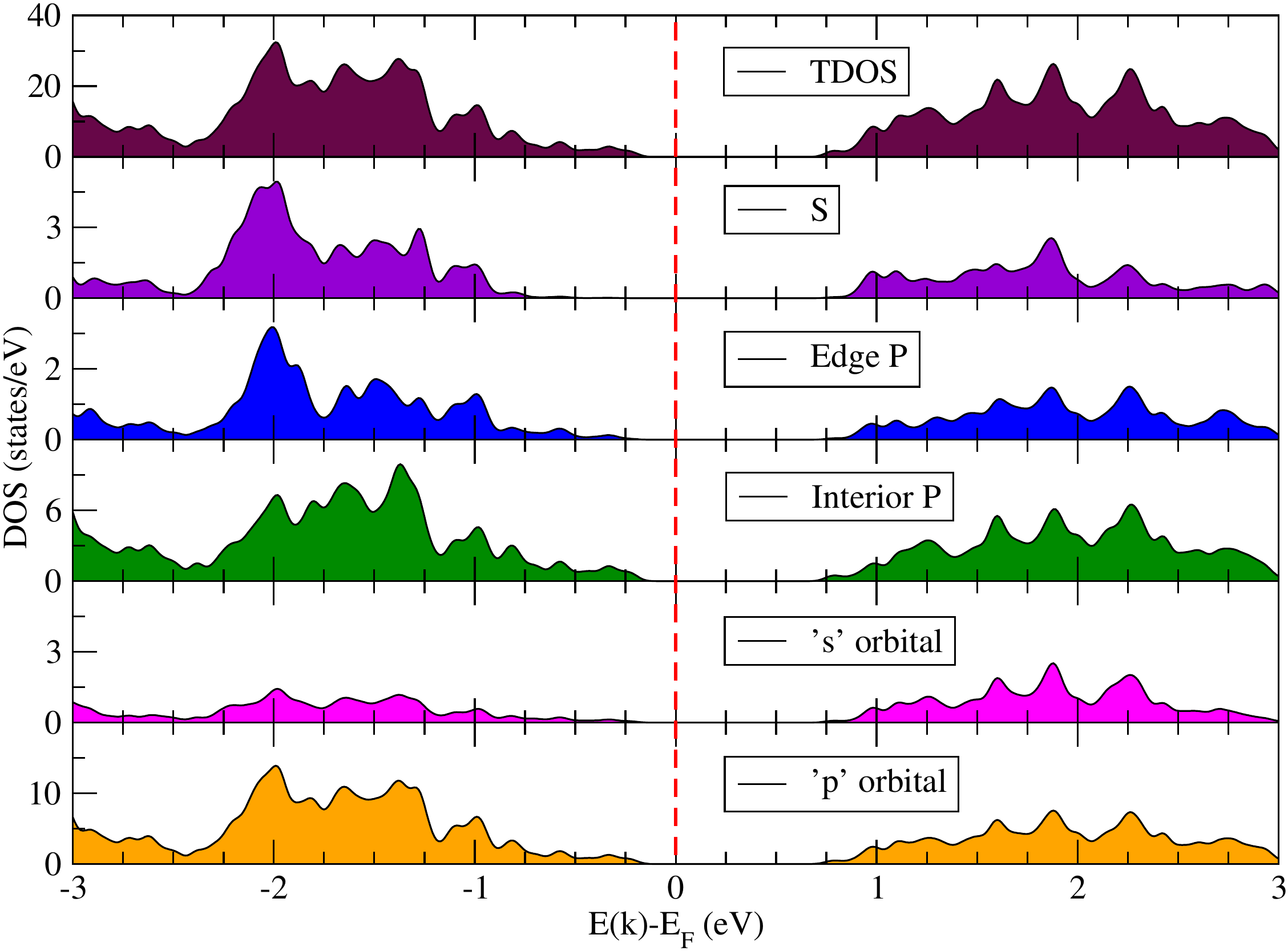}

}~~~~~~~~~\subfloat[Se]{\includegraphics[scale=0.25]{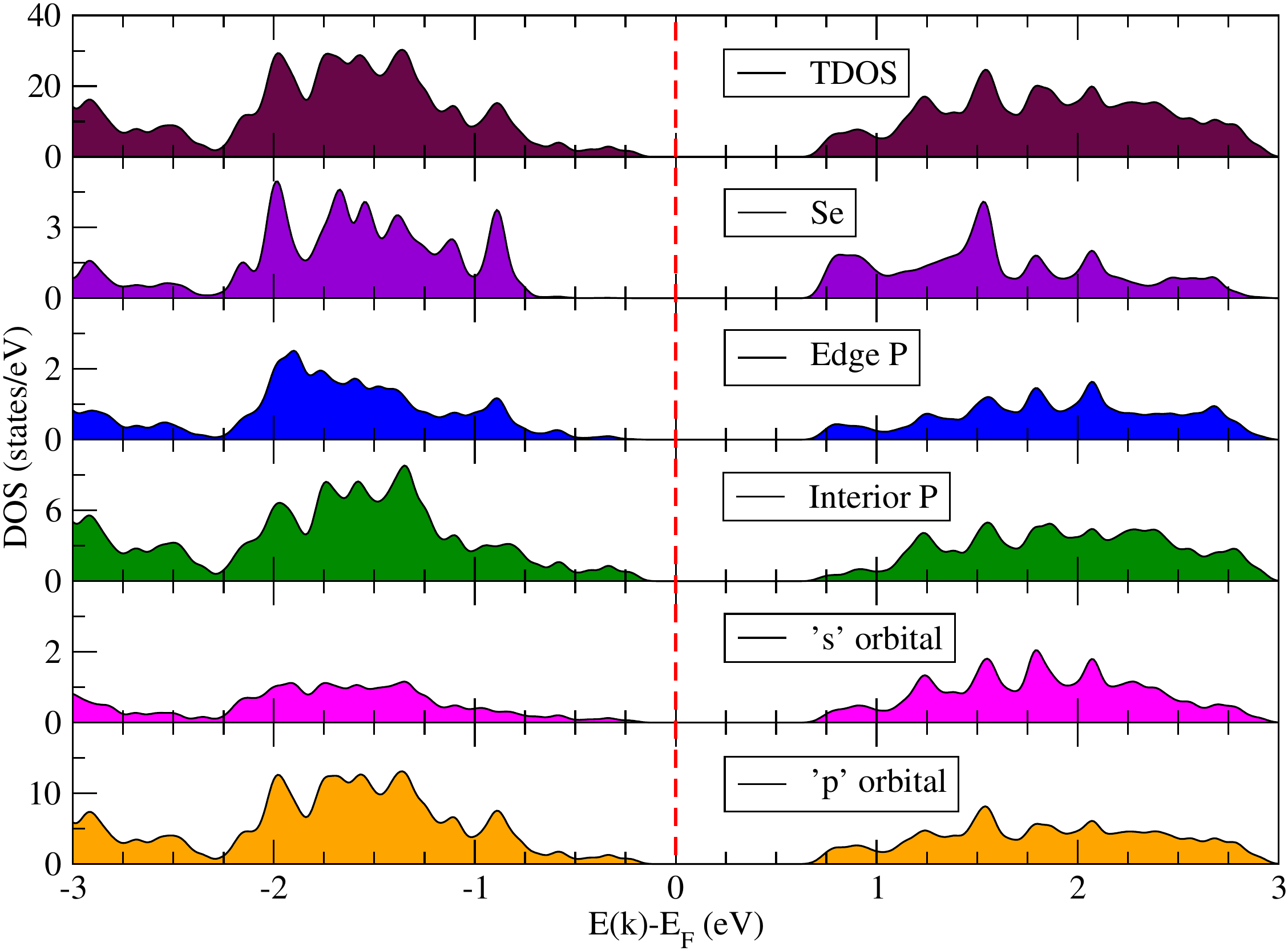}

}
\par\end{centering}
\caption{Total, atom projected, and orbital projected density of states (PDOS)
of pristine and saturated 11-APNRs. The atom projected DOS is further
classified based on the contributions from the passivating groups,
as well as from P atoms located on the edges (``Edge P''), and the
interior (``Interior P'') of the APNRs. \label{fig:DOS}}
\end{figure}

The total, orbital projected, and atom projected density of states
(DOS) for the case of 11-APNRs are presented in Fig. \ref{fig:DOS}.
For the pristine ribbon, we note that the edge P atoms contribute
more to the conduction band states near the Fermi level, as compared
to the interior P atoms; while for the valence bands, both the interior
as well edge atoms make significant contributions. For the passivated
APNRs, we see that both the edge and the interior P atoms make significant
contributions to the frontier bands. Furthermore, increasing contribution
of the passivating atoms to the bands near the Fermi level, with their
increasing sizes is also obvious from various DOS plots. It is also
clear from these plots that the p-type orbitals contribute more than
the s-type orbitals to the frontier states.

\subsection{Optical absorption spectra}

Based upon the single-particle DFT-PBE level band structure calculations,
we computed the optical absorption spectra of pristine, as well as
edge-saturated N-APNRs with various functional groups, for N=3--24,
with the incident light polarized along the length of the ribbons.
The spectrum was calculated according to the standard formalism, by
computing the imaginary part of the dielectric constant matrix, as
implemented in the VASP program.\citep{Kresse_et_al,Kresse_2_et_al}

\begin{figure}[h]
\begin{centering}
\includegraphics[scale=0.45]{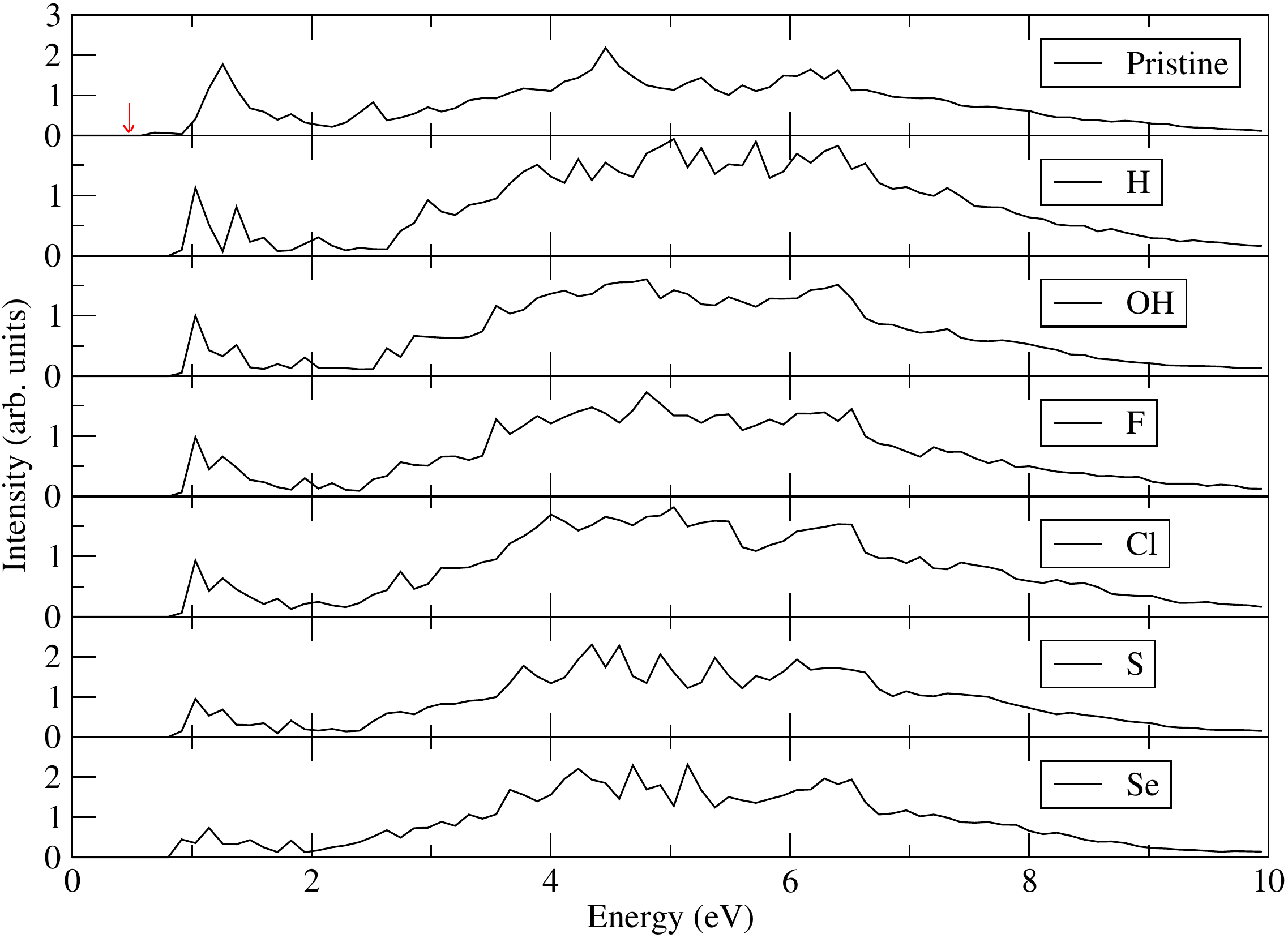}
\par\end{centering}
\caption{Single-particle optical absorption spectra of bare and edge-saturated
11-APNRs computed using DFT-PBE based formalism. The calculated band-gap
is indicated by a red arrow for the pristine structure. In other cases
the first peak position represents the band gap energy.\label{fig:optical}}
\end{figure}

In Fig. \ref{fig:optical} we present the calculated optical absorption
spectra of pristine and edge-passivated 11-APNRs. To the best of our
knowledge, the only previous calculation of the absorption spectrum
of APNRs, based upon the first-principles DFT methodology, was reported
by Tran \emph{et al.}\citep{Tran_et_al}, for H-passivated ribbons,
but only up to 2.4 eV. Here we report calculations not just on H-passivated
APNRs, but also on pristine ribbons, and also those passivated by
other groups. In Fig. S3 of Supporting Information, we also present
the results of similar calculations on narrower 5-APNRs, and broader
24-APNRs.

To investigate the effects of spin-orbit coupling (SOC) on the absorption
spectra of APNRs, we carried out calculations for hydrogen passivated
5- and 11-APNRs, and in Fig. S6 of the Supporting Information, we
compare the spectra computed with and without SOC. For the narrower
ribbon, i.e., 5-APNR, no significant change in peak locations as well
as intensity was observed. But, for the broader ribbon, i.e., 11-APNR,
although no noticeable change in the peak locations is seen, but the
peak intensities do get weaker for energies beyond 5 eV. 

In order to benchmark the calculated spectra, in Table \ref{tab:Comparing-peak}
we compare the locations of our absorption peaks, with those reported
by Tran \emph{et al.}\citep{Tran_et_al}, for H-passivated ribbons.
Because Tran \emph{et al.}\citep{Tran_et_al} reported the spectra
only up to 2.4 eV, therefore, for N=3 and N=5, we are able to compare
only one peak each, while for N=10, comparison for three peaks is
possible (see Table \ref{tab:Comparing-peak}). We note that the our
peak locations are about 0.1 eV blue shifted as compared to the ones
reported by Tran \emph{et al.}\citep{Tran_et_al}, which is a fairly
good agreement given the fact that they employed a different computer
code, Quantum Espresso\citep{quantum-espresso}, for their calculations.

\begin{table}
\caption{Comparing the peak positions (in eV) in optical absorption spectra
of hydrogen-saturated N-APNRs computed by us, with those reported
by Tran and Yang\citep{Tran_et_al}. For N=10, first three peak locations
are compared. Both the calculations are based on DFT-PBE formalism.\label{tab:Comparing-peak}}

\centering{}%
\begin{tabular}{ccccc}
\hline 
N & $\qquad$ & This work & $\qquad$ & Tran et al.(Ref.\citep{Tran_et_al})\tabularnewline
\hline 
\hline 
3 &  & 2.05 &  & 1.93\tabularnewline
 &  &  &  & \tabularnewline
5 &  & 1.37 &  & 1.25\tabularnewline
 &  &  &  & \tabularnewline
10 &  & %
\begin{tabular}{c}
1.03\tabularnewline
1.49\tabularnewline
2.29\tabularnewline
\end{tabular} &  & %
\begin{tabular}{c}
0.93\tabularnewline
1.4\tabularnewline
2.19\tabularnewline
\end{tabular}\tabularnewline
\hline 
\end{tabular}
\end{table}

On examining the absorption spectra of 11-APNRs in Fig. \ref{fig:optical},
we see the following trends: (a) for the pristine ribbon there is
no peak at the location of its band gap, consistent with the fact
that it is an indirect band gap semiconductor, (b) for all the passivated
ribbons, as expected, the first peak occurs at the location of the
band gap, and (c) for all the APNRs, the highest intensity peak occurs
at energies larger than 4 eV.

We present the locations of the first, the second, and the most intense
peaks of optical absorption spectra of pristine and passivated 11-APNRs
in the Table \ref{tab:peak-location}. We also investigated the orbitals
involved in the one-electron transitions leading to these peaks, and
they are also presented in the same Table, in parentheses next to
the peak energies. We note that for the pristine APNR, the first peak
occurs at 0.69 eV, caused by the orbital transition $|V\rightarrow C+1\rangle$,
for a nonzero momentum value, very close to the $\Gamma$ point. For
all the passivated APNRs, the first peak appears due to the $\arrowvert V\rightarrow C\rangle$
transition, occurring at the $\Gamma$ point. The second peak for
the pristine ribbon is due to the transition $\arrowvert V\rightarrow C+2\rangle$,
slightly away from the $\Gamma$ point. For APNRs passivated by monovalent
groups, namely, H, OH, F and Cl, transition $\arrowvert V\rightarrow C+3\rangle$
gives rise to the second peak, while the same peak for the divalent
passivating groups S and Se, is due to transitions $\arrowvert V-1\rightarrow C+2\rangle$,
and $\arrowvert V\rightarrow C+2\rangle$, respectively. In all the
cases, these optical transitions occur either at the $\Gamma$ point,
or fairly close to it. Finally, we examine the nature of the highest
intensity peaks located at higher energies. Quite expectedly, the
corresponding transitions involves bands far away from the Fermi level
(see Table \ref{tab:peak-location}), and they occur closer to the
edge of the Brillouin zone, as shown for the case of H-saturated 11-APNR,
in Fig. S2 of the Supporting Information. 
\begin{table}
\caption{Locations of the first, second, and the most intense peaks, and the
bands involved in the transition (in the parentheses), in the optical
absorption spectra of pristine and saturated 11-APNRs. Below $V$
and $C$ denote the highest valance band, and the lowest conduction
band, respectively. Similarly $V-n$ ($C+n)$ denotes the $n$-th
valence (conduction) band, counting away from the Fermi level.\label{tab:peak-location}}

\begin{tabular}{|c|c|c|c|}
\hline 
Saturation & First peak & Second peak & Most intense peak\tabularnewline
\hline 
 & position (eV) & position (eV) & position (eV)\tabularnewline
\hline 
\hline 
Pristine & 0.69 ($\arrowvert V\rightarrow C+1\rangle$) & 1.26 ($\arrowvert V\rightarrow C+2\rangle$) & 4.46 ($\arrowvert V-8\rightarrow C+9\rangle$)\tabularnewline
\hline 
H & 1.03 ($\arrowvert V\rightarrow C\rangle$) & 1.38 ($\arrowvert V\rightarrow C+3\rangle$) & 5.04 ($\arrowvert V-10\rightarrow C+10\rangle$)\tabularnewline
\hline 
OH & 1.03 ($\arrowvert V\rightarrow C\rangle$) & 1.38 ($\arrowvert V\rightarrow C+3\rangle$) & 4.81 ($\arrowvert V-10\rightarrow C+8\rangle$)\tabularnewline
\hline 
F & 1.03 ($\arrowvert V\rightarrow C\rangle$) & 1.28 ($\arrowvert V\rightarrow C+3\rangle$) & 4.82 ($\arrowvert V-10\rightarrow C+9\rangle$)\tabularnewline
\hline 
Cl & 1.03 ($\arrowvert V\rightarrow C\rangle$) & 1.27 ($\arrowvert V\rightarrow C+3\rangle$) & 5.05 ($\arrowvert V-11\rightarrow C+12\rangle$)\tabularnewline
\hline 
S & 1.03 ($\arrowvert V\rightarrow C\rangle$) & 1.27 ($\arrowvert V-1\rightarrow C+2\rangle$) & 4.35 ($\arrowvert V-8\rightarrow C+8\rangle$)\tabularnewline
\hline 
Se & 0.92 ($\arrowvert V\rightarrow C\rangle$) & 1.14 ($\arrowvert V\rightarrow C+2\rangle$) & 5.16 ($\arrowvert V-14\rightarrow C+13\rangle$)\tabularnewline
\hline 
\end{tabular}
\end{table}

At this point, one may wonder as to how close are the absorption spectra
of our widest system 24-APNR to that of the infinite phosphorene monolayer.
To investigate that, in Fig. S5 of the Supporting Information, we
compare the optical absorption spectra of H-passivated 24-APNR to
that of monolayer phosphorene, both computed using the DFT-PBE approach.
From the figure it is obvious that the two spectra are very similar
to each other in terms of intensity profile, as well as peak locations,
except that the monolayer spectrum, quite expectedly, is much more
intense. Thus, for all practical purposes we can assume that the optical
properties of 24-APNR have saturated to the monolayer value.

When we compare the absorption spectra of 11-APNRs with those of 5-APNRs
and 24-APNRs (see Fig. S3, Supporting Information) we note that for
the pristine 5-APNR, the first absorption peak is at the band gap,
because it is a direct band gap material. For the passivated nanoribbons,
we note that the basic qualitative features of the absorption spectra
are the same irrespective of the widths. For narrower ribbons the
absorption peaks are sharper and well separated, while for broader
ones they evolve into absorption bands. 

We also note that the qualitative features of the absorption spectrum
of H-passivated 5-APNR computed using the GW+BSE approach\citep{Nourbakhsh_et_al}
are quite similar to that computed by us using the PBE-DFT approach,
except, of course, for the peak locations.

\section{Conclusions}

\label{sec:conclusions}

In this work, we presented results on first-principles DFT calculations
on pristine and passivated $N$-APNRs, ranging from the very narrow
($N=3$), to the very broad ($N=24$). We first performed geometry
optimization for each ribbon, and for those geometries computed quantities
such as the formation energies, the band gaps, the band structure,
and the optical absorption spectra, using the PBE functional. In addition,
for a selected few ribbons, we also calculated the band gaps using
the HSE06 functional, and found that it yields band gaps significantly
larger than those predicted by PBE-functional based calculations.
This implies that electron correlations make important contributions,
highlighting their importance in the reduced dimensional systems such
as APNRs. Therefore, by reducing the dimension of a system, one can
manipulate the band gap as desired for a particular device application.
Lower-dimensional materials can be useful in many applications such
as sensors, information storage, optoelectronic devices, and transport
and spintronic applications. 

According to formation energy calculations, the pristine APNRs were
predicted to be unstable, however, the results may change once the
electron-correlation effects are taken into account.\textcolor{red}{{}
}Formation energies also suggest that the narrower ribbons are more
favorable than the wider ones, and that F- and OH-passivated ribbons
are stabler as compared to other passivated ribbons.

Our calculations predict all pristine APNRs to be indirect band gap
semiconductors, except for 5-APNR which was shown to have a direct
band gap. Irrespective of the passivating group, all edge-saturated
APNRs were found to be direct band gap materials, with the gap located
at the $\Gamma$ point. With the increasing width, band gaps of the
passivated nanoribbons were shown to evolve to the band gaps of infinite
monolayer phosphorene. However, pristine nanoribbons saturated to
much smaller band gaps with increasing widths, indicating that the
dangling bonds, and the related edge reconstruction, play important
roles in their electronic properties. 

To examine the influence of the relativistic effects, we also performed
calculations on a couple of H-passivated APNRs including the spin-orbit
coupling, and found no significant changes either in the band structure,
or in the intensity profiles of the absorption spectra.

In this work we also presented a first-time systematic study of the
dependence of optical absorption spectra on the passivating groups.
We found that the first absorption peak corresponds to $V\rightarrow C$
excitation at the $\Gamma$ point, corresponding to the band gaps,
irrespective of the group. We also analyzed the bands involved in
the higher energy transitions. As the self-energy corrections and
excitonic effects were not incorporated in our calculations, detailed
prediction of absorption profiles, which can be directly compared
with the experiments, is not possible. It will, therefore, be interesting,
in future, to perform calculations based upon GW-approximation and
Bethe-Salpeter equations, to account for the influence of electron-correlation
effects on the band structure and optical properties of wider APNRs.

\section*{Author Information }

\subsection*{Corresponding Authors}

Alok Shukla:  {*}E-mail: shukla@phy.iitb.ac.in

\subsection*{Notes}

The authors declare no competing financial interests.

\section*{Acknowledgements}

Work of P.B. was supported by a Senior Research Fellowship offered
by University Grants Commission, India. Also, we are thankful to the
space-time server of IIT Bombay, India for providing computational
facility to perform the calculations.

\bibliographystyle{apsrev4-1}
\addcontentsline{toc}{section}{\refname}\bibliography{apnr}

\end{document}

\title{Supporting Information\\
Influence of Edge Functionalization on Electronic and Optical Properties
of Armchair Phosphorene Nanoribbons: a First-Principles Study}
\author{Pritam Bhattacharyya}
\email{pritambhattacharyya01@gmail.com}

\address{Department of Physics, Indian Institute of Technology Bombay, Powai,
Mumbai 400076, India}
\author{Rupesh Chaudhari}
\affiliation{Department of Physics, Indian Institute of Technology Bombay, Powai,
Mumbai 400076, India}
\author{Naresh Alaal}
\email{nareshkdnr@gmail.com}

\affiliation{Physical Sciences and Engineering Division, King Abdullah University
of Science and Technology, Thuwal 23955-6900, Saudi Arabia}
\author{Tushar Rana}
\affiliation{Department of Physics and Nanotechnology, SRM Institute of Science
and Technology,SRM Nagar, Kattankulathur - 603203 (Tamil Nadu), India}
\author{Alok Shukla}
\email{shukla@phy.iitb.ac.in}

\affiliation{Department of Physics, Indian Institute of Technology Bombay, Powai,
Mumbai 400076, India}

\maketitle
\begin{table}
\caption{Total energies of the molecules computed using the PBE functional
used for computing the formation energies of passivated APNRs}

\centering{}%
\begin{tabular}{ccc}
\hline 
Molecule & \hspace{0.5cm} & Energy (eV)\tabularnewline
\hline 
\hline 
H\textsubscript{2} &  & -6.7009\tabularnewline
O\textsubscript{2} &  & -9.8611\tabularnewline
F\textsubscript{2} &  & -3.5524\tabularnewline
Cl\textsubscript{2} &  & -3.5559\tabularnewline
S\textsubscript{2} &  & -6.5951\tabularnewline
Se\textsubscript{2} &  & -5.4183\tabularnewline
\hline 
\end{tabular}
\end{table}

\begin{figure}[h]
\subfloat[Top view]{\includegraphics[scale=0.3]{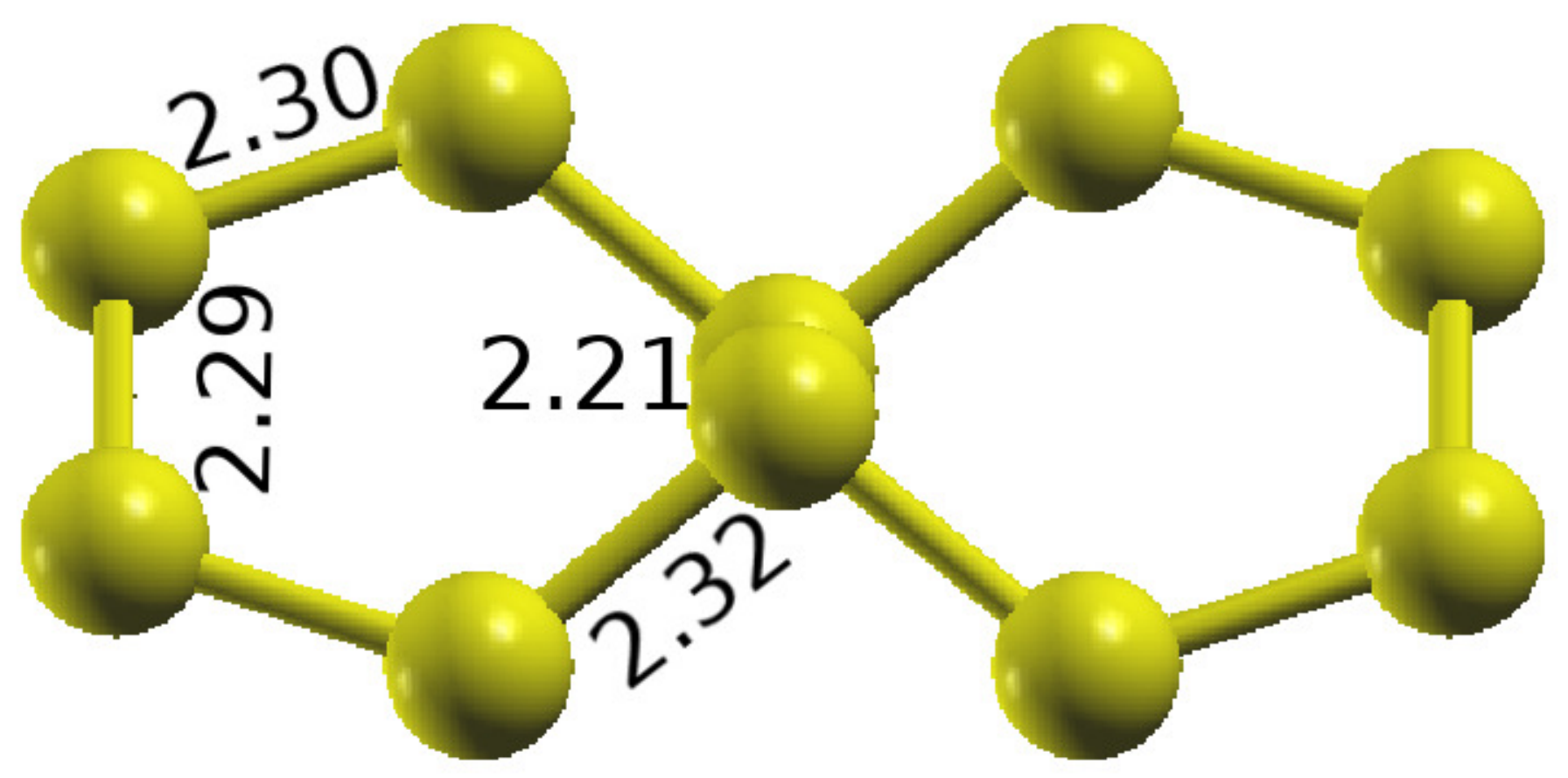}

}~~~~~~~~~~\subfloat[Tilted side view]{\includegraphics[scale=0.25]{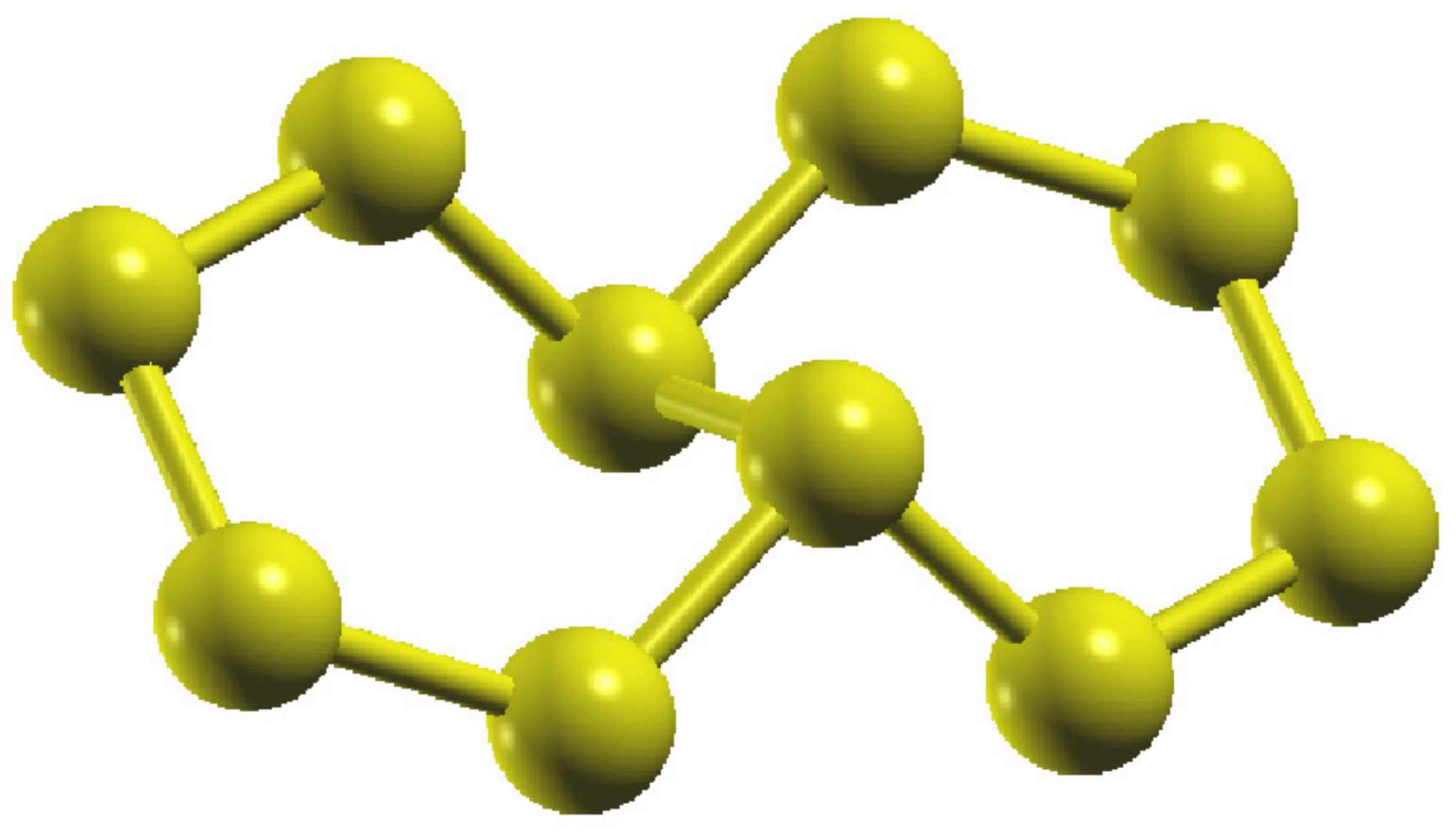}

}

\caption{Optimized geometry of pristine 5-APNR}

\end{figure}

\begin{figure}[h]
\includegraphics[scale=0.4]{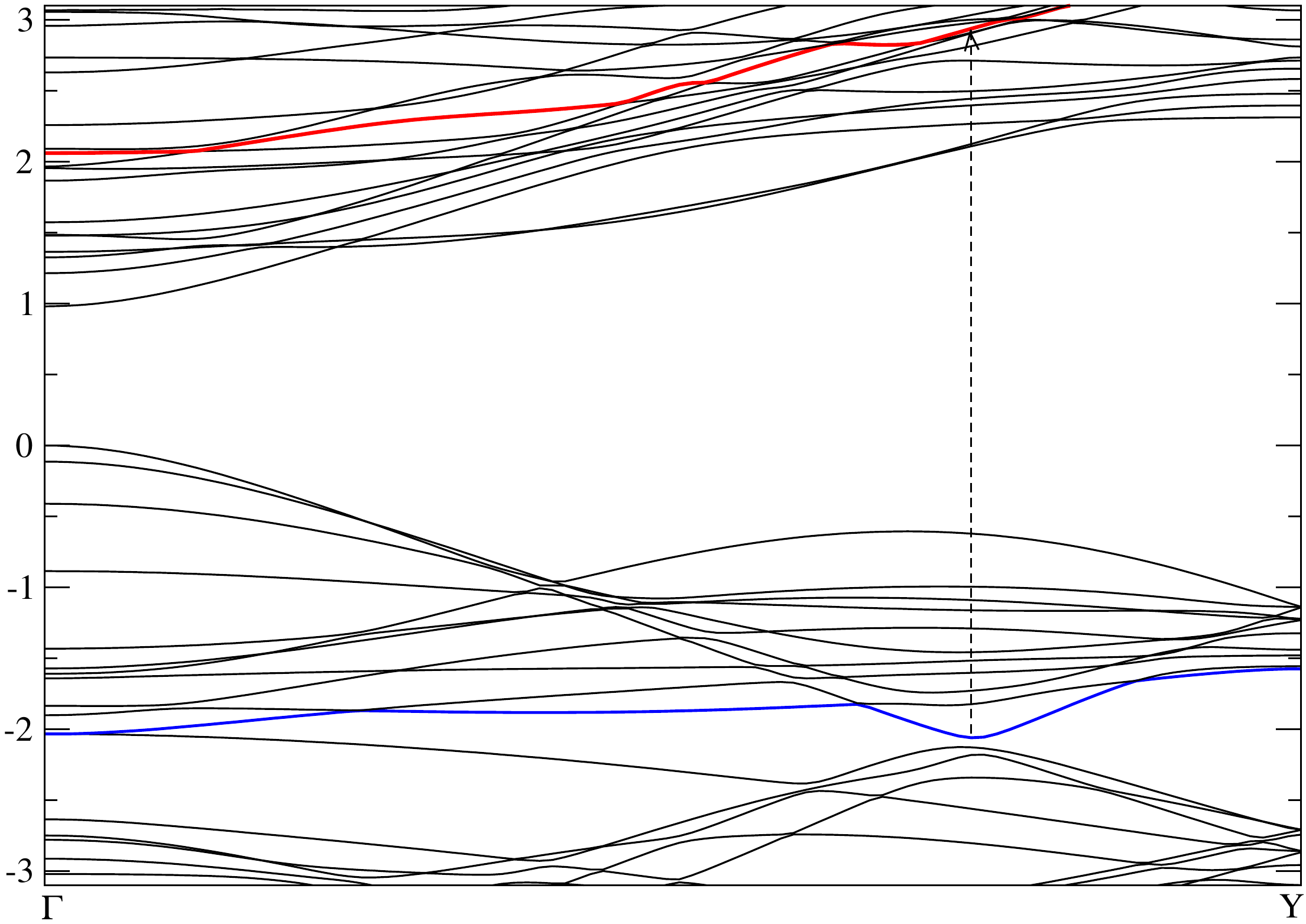}

\caption{The arrow indicates the $k$ value, as well as the involved bands,
corresponding to the direct transition responsible for the most intense
peak in the optical absorption spectrum of H-saturated 11-APNR. The
involved valence and conduction bands are in blue and red colors,
respectively.}
\end{figure}

\begin{figure}
\subfloat[5-APNR]{\includegraphics[scale=0.4]{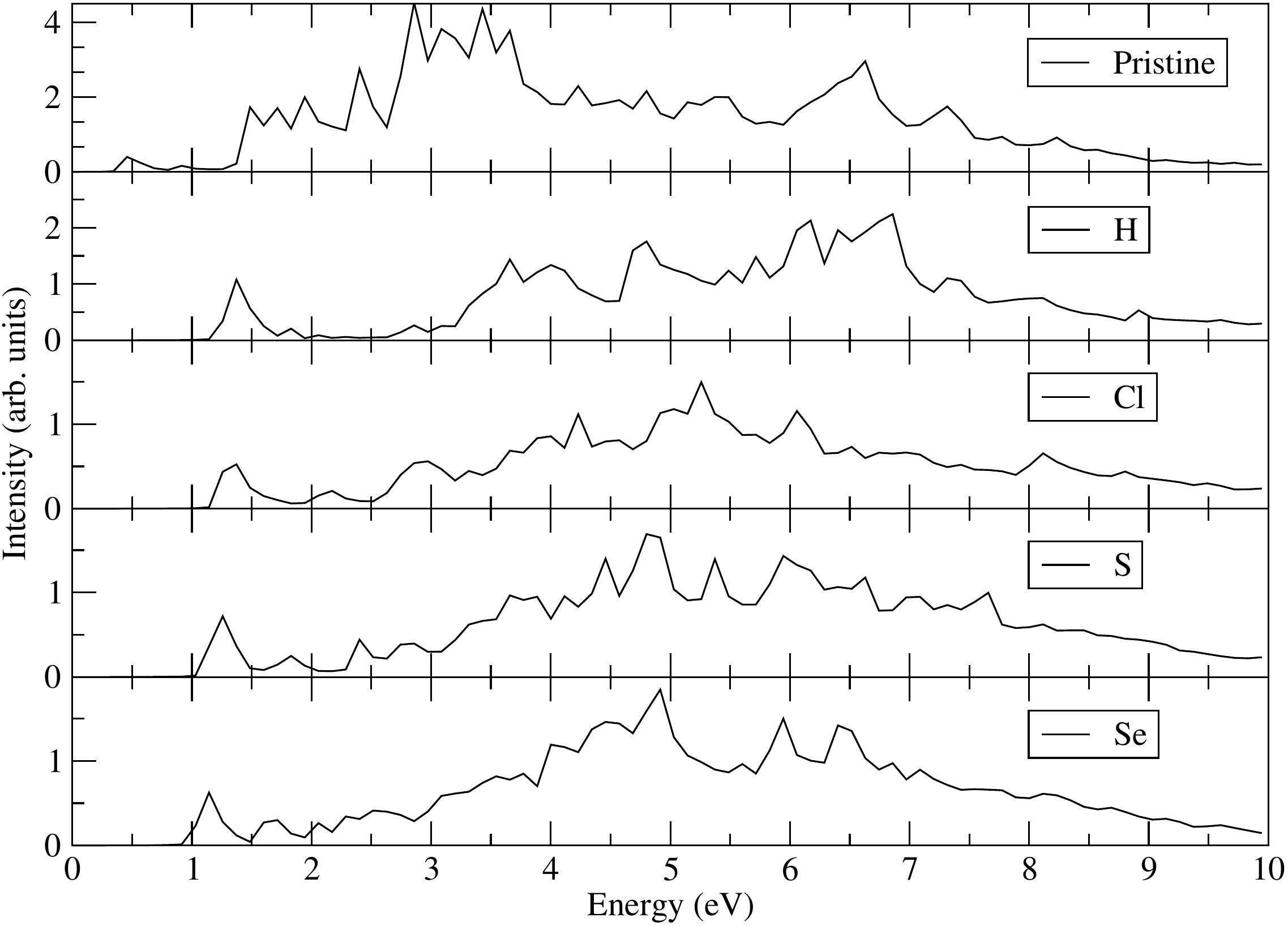}

}\subfloat[24-APNR]{\includegraphics[scale=0.4]{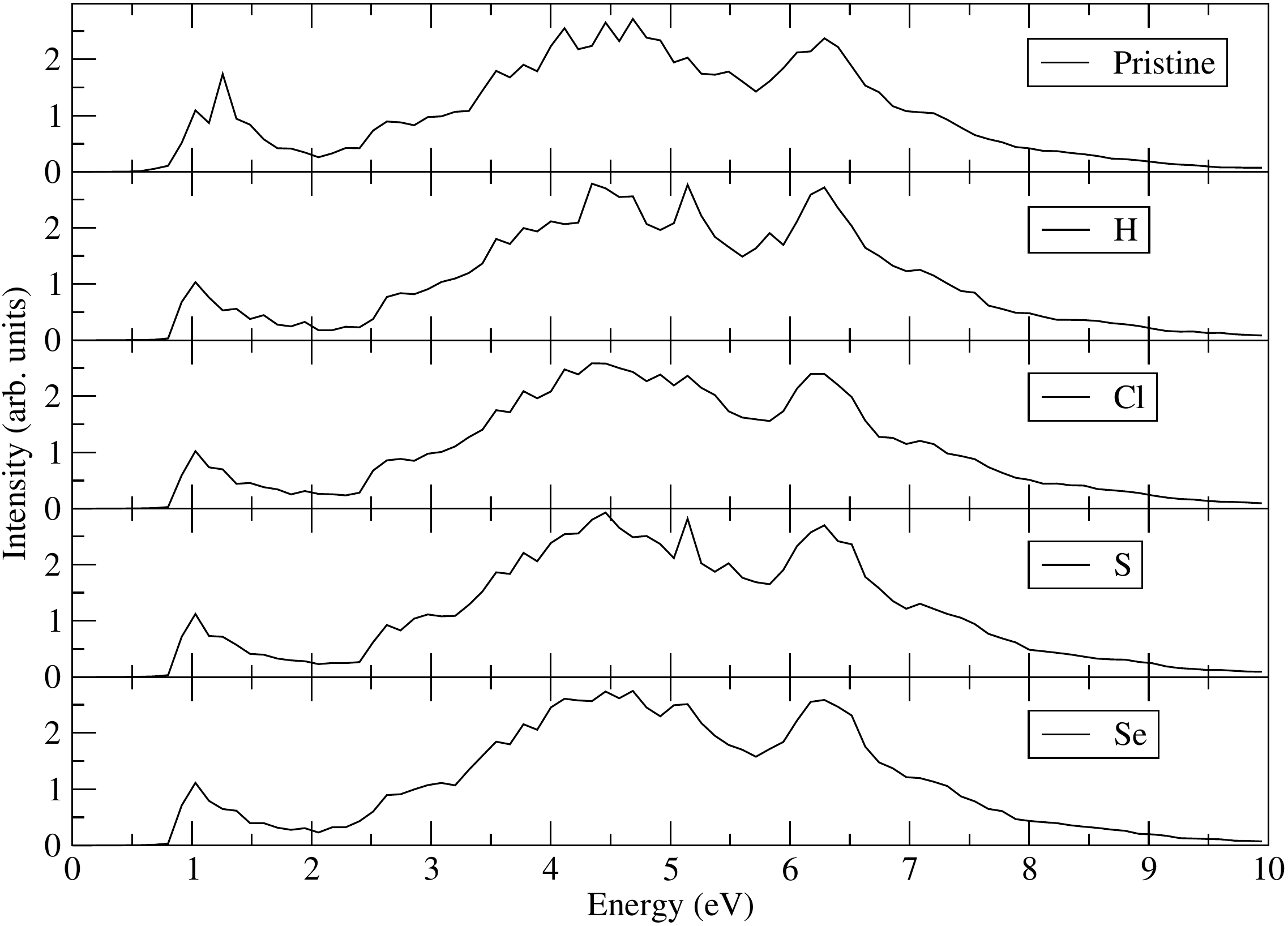}

}

\caption{Optical absorption spectra of bare and saturated 5- and 24-APNR computed
using the DFT-PBE methodology. In the pristine case of 5-APNR, the
first peak of the optical spectrum is due to the transition at the
band edge.}

\end{figure}

\begin{figure}
\includegraphics[scale=0.5]{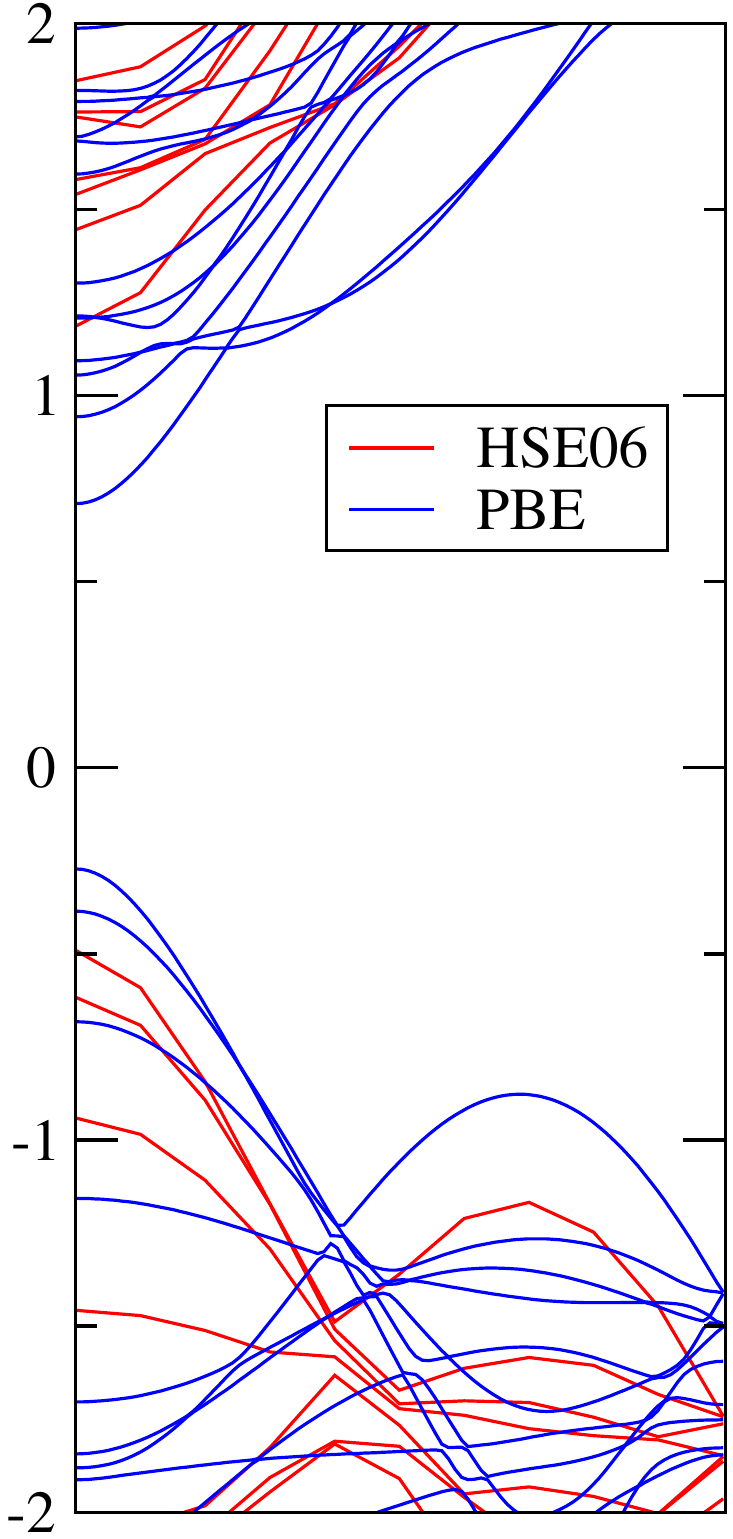}

\caption{Comparison of band structures of hydrogen(H)-saturated 11-APNR, computed
using the PBE and HSE06 hybrid functionals.}
\end{figure}

\begin{figure}
\includegraphics[scale=0.4]{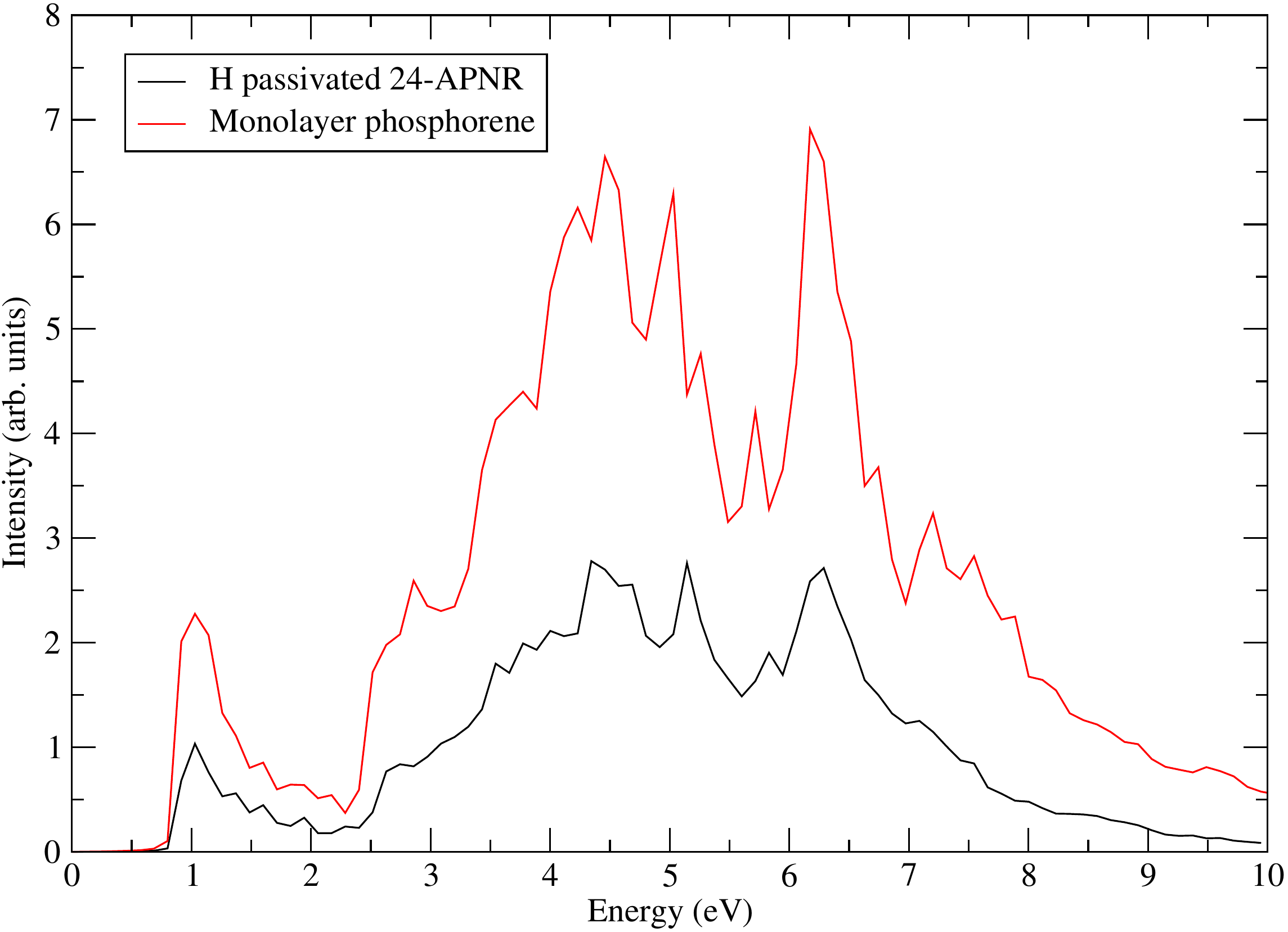}

\caption{Optical absorption spectra of monolayer phosphorene compared to that
of hydrogen-passivated 24-APNR, computed using the DFT-PBE methodology.
Two spectra are very similar to each other except for larger intensity
for the infinite monolayer.}
\end{figure}

\begin{figure}
\subfloat[5-APNR]{\includegraphics[scale=0.35]{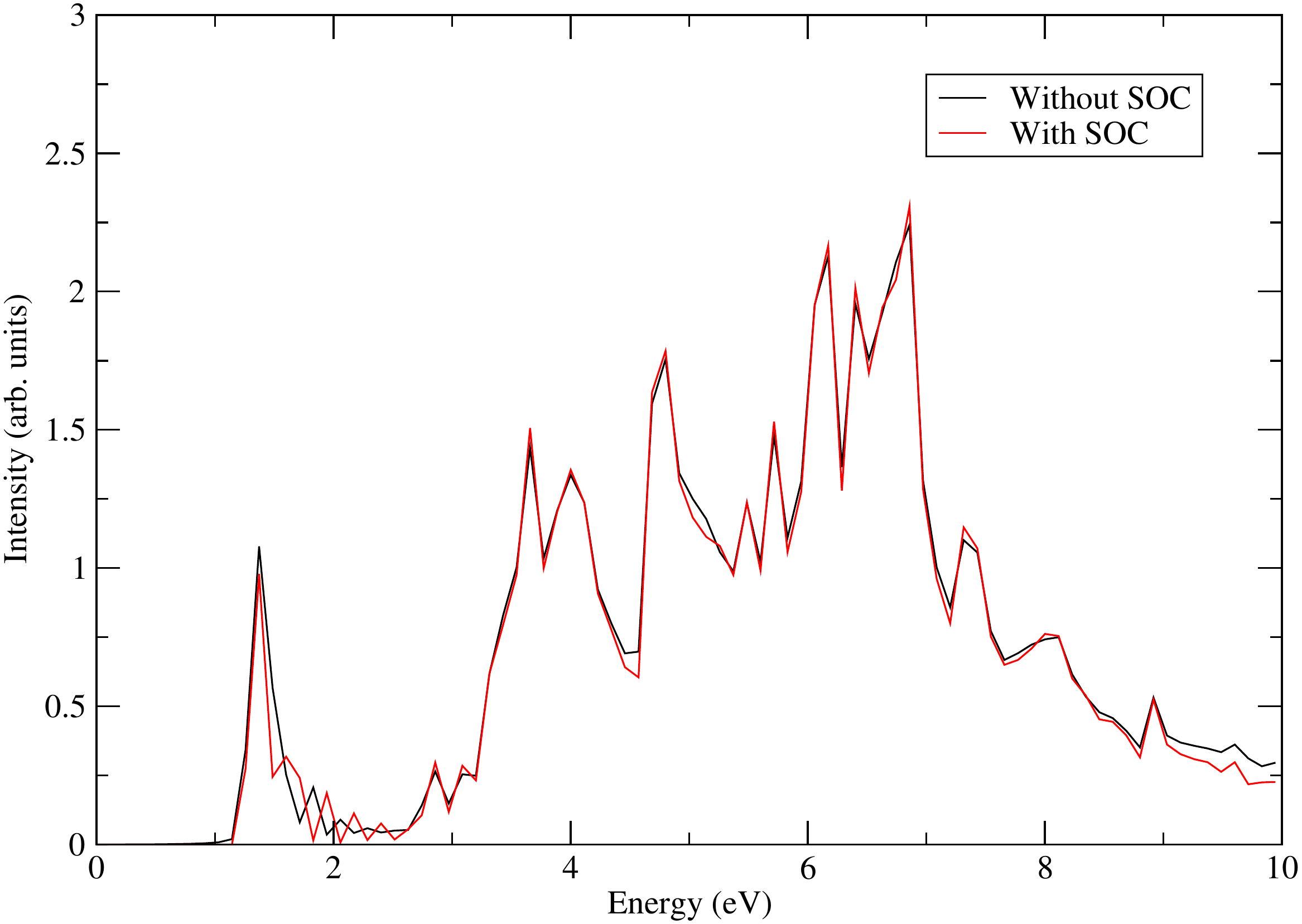}

}\subfloat[11-APNR]{\includegraphics[scale=0.35]{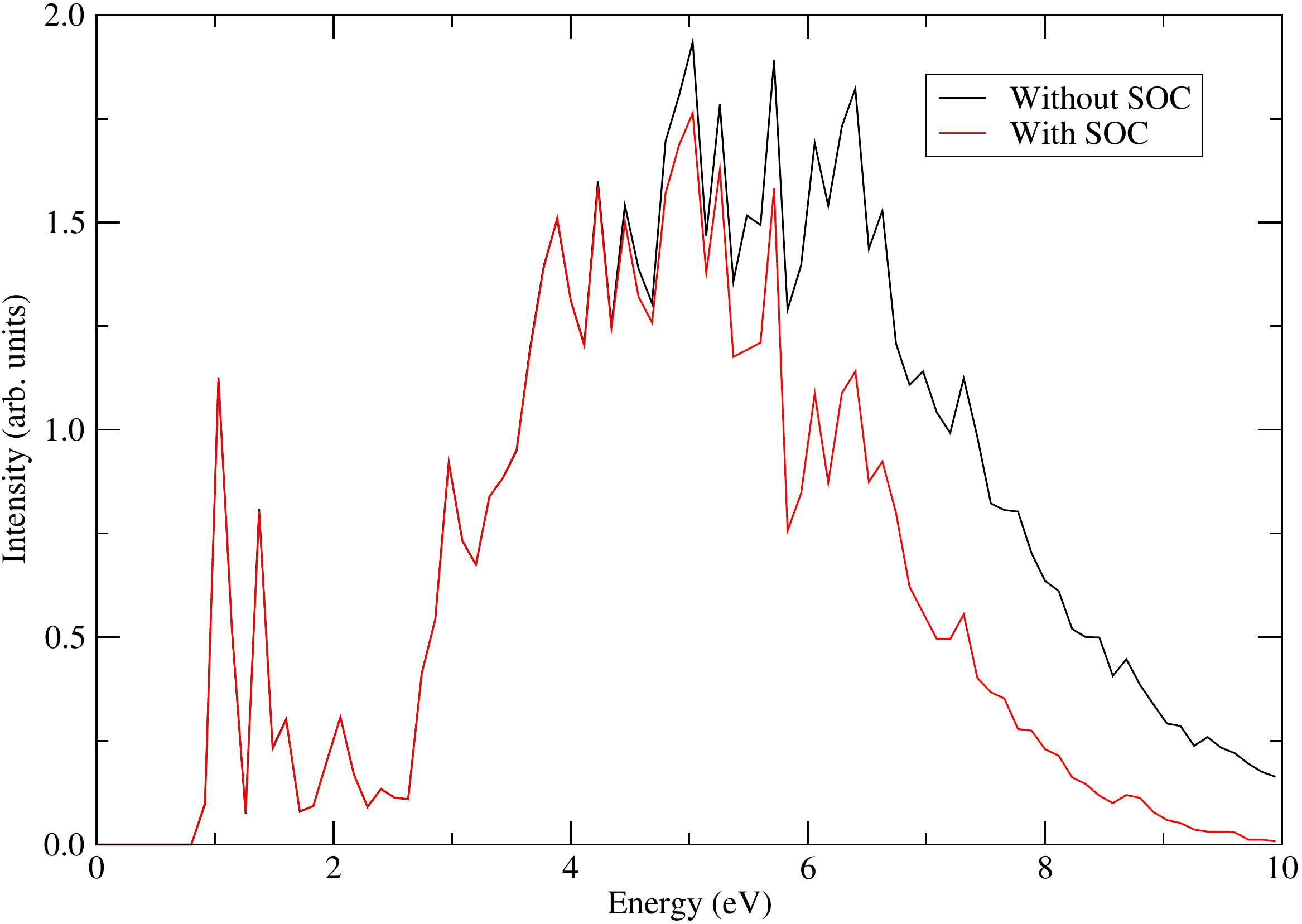}

}

\caption{Optical absorption spectra of hydrogen-passivated (a) 5-APNR and (b)
11-APNR, without and with including the spin-orbit coupling (SOC),
computed using the DFT-PBE methodology. }
\end{figure}